\documentclass[11pt]{article}
\usepackage{hyperref}
\usepackage{graphicx}
\usepackage{graphics}
\usepackage{dsfont}
\usepackage{epsfig}
\usepackage{amsmath,amssymb,amsthm,amscd}

\newcommand{\tr}{\textrm{tr}}
\newcommand{\Res}{\textrm{Res}}

\numberwithin{equation}{section}

\begin{document}
\begin{titlepage}
{}~ \hfill\vbox{ \hbox{} }\break

\rightline{IPMU 12-0093}

\vskip 3 cm

\centerline{\Large \bf  On Gauge Theory and Topological String}    \vskip 0.5 cm
 \centerline{\Large \bf   in Nekrasov-Shatashvili Limit}    \vskip 0.5 cm
\renewcommand{\thefootnote}{\fnsymbol{footnote}}
\vskip 30pt \centerline{ {\large \rm Min-xin Huang
\footnote{minxin.huang@ipmu.jp}    } } \vskip .5cm \vskip 30pt

\begin{center}
{Kavli Institute for the Physics and Mathematics of the Universe (Kavli IPMU),  \\ \vskip 0.2 cm  University of Tokyo, Kashiwa, Chiba 277-8582, Japan}\\ [3 mm]
\end{center}

\setcounter{footnote}{0}
\renewcommand{\thefootnote}{\arabic{footnote}}
\vskip 60pt
\begin{abstract}
We study the Nekrasov-Shatashvili limit of the $\mathcal{N}=2$ supersymmetric gauge theory and topological string theory on certain local toric Calabi-Yau manifolds. In this limit one of the two deformation parameters $\epsilon_{1,2}$ of the $\Omega$ background is set to zero and we study the perturbative expansion of the topological amplitudes around the remaining parameter. We derive differential equations from Seiberg-Witten curves and mirror geometries, which determine the higher genus topological amplitudes up to a constant. We show that the higher genus formulae previously obtained from holomorphic anomaly equations and boundary conditions satisfy these differential equations. We also provide a derivation of the holomorphic anomaly equations in the  Nekrasov-Shatashvili limit from these differential equations.

\end{abstract}

\end{titlepage}
\vfill \eject


\newpage

\baselineskip=16pt

\tableofcontents

\section{Introduction}

There has been much progress in the study of supersymmetric gauge theories since Seiberg and Witten discovered that the $\mathcal{N}=2$ supersymmetric gauge theories are exactly solvable \cite{SW1, SW2}. The prepotential which characterizes the effective action can be determined by holomorphicity and monodromy in the moduli space. On the other hand, the instanton contributions in the prepotential can be directly computed by Nekrasov partition function \cite{Nekrasov}. The Nekrasov partition function is parametrized by two parameters $\epsilon_1$ and $\epsilon_2$ which deform the $\mathbb{R}^4$ space. It can be shown by saddle point method that the leading order contribution of Nekrasov function in small $\epsilon_1, \epsilon_2$ is equal to the Seiberg-Witten prepotential \cite{NO}. For more mathematical perspectives of the Nekrasov  function see e.g. \cite{Braverman, NY}. Furthermore, the higher order contributions in $\epsilon_1, \epsilon_2$ expansion of the Nekrasov function compute the gravitational coupling terms in the effective action, and is analogous to the higher genus amplitudes in topological string theory which can be computed by the method of holomorphic anomaly equation \cite{BCOV} and gap conditions in the moduli space proposed in \cite{HK2006, HK2009}. The two parameters $\epsilon_1, \epsilon_2$ correspond to a refinement of the string coupling in topological string theory, which was studied for certain toric Calabi-Yau manifolds in \cite{IKV}. The homomorphic anomaly equation and gap conditions can be extended to the refined case and the higher order terms in $SU(2)$ Nekrasov function are solved exactly \cite{HK2010, HKK, KW}. The higher genus formulae are expressed in terms of quasi-modular forms such as Eisenstein series and Jacobi theta functions, and the formulae are exact in the sense that they sum up all instanton contributions at a fixed genus.  

The Nekrasov partition function can be also related to the correlation function of 2d Liouville theory by the AGT (Alday-Gaiotto-Tachikawa) conjecture \cite{AGT}. Recently there have been many works in this direction. We hope our works can provide some ideas for the AGT conjecture. 

We will consider the so called Nekrasov-Shatashvili limit, also sometimes known as the chiral limit,  of the Nekrasov function, which sets one of the deformation parameter $\epsilon_2=0$ and we expand the Nekrasov function for small $\epsilon \equiv \epsilon_1$. Nekrasov and Shatashvilli conjectures in this limit the $\mathcal{N}=2$ gauge theories are described by certain quantum integrable systems \cite{NS}. The quantum integrable systems provides another way to compute the Nekrasov function in the $\epsilon_2=0$ limit and has been considered in e.g. \cite{MM, ACDKV}. In our previous paper \cite{HKK} we showed that the formulae we derived from holomorphic anomaly equation satisfy the quantum equations in the sine-Gordon model for the pure $SU(2)$ Seiberg-Witten theory. Thus, if the quantum integrable system description of the Nekrasov-Shatashvili limit is correct, our higher genus formulae in this limit would be exactly proven. 

In this paper we study the approach of using the saddle point method to compute Nekrasov function in the Nekrasov-Shatashvili limit. This is carried out quite explicitly in the papers \cite{FMPP, Poghossian}, and seems to be on a more solid footing than the approach of using quantum integrable systems mentioned above. Furthermore, the saddle point method is readily applicable to the case of Seiberg-Witten gauge theory with matters and to higher rank gauge group. We will show that our $SU(2)$ higher genus formulae \cite{HK2010,HKK} in the Nekrasov-Shatashvili limit satisfy the saddle point equations in \cite{FMPP, Poghossian}. Since these equations uniquely fix the higher genus contributions (up to some constants, which can be easily checked),  we would have proven our formulae exactly.

\section{Review of the saddle point method}

We will be interested in the small $\epsilon$ expansion of the logarithm of the Nekrasov partition function, which is  called the free energy
\begin{eqnarray} \label{1111.2.1}
\log Z (\epsilon_1,\epsilon_2, a_i) =\sum_{g,n=0}^{\infty}  (\epsilon_1+\epsilon_2)^{2n}(\epsilon_1\epsilon_2)^{g-1} F^{(n,g)}(a_i)
\end{eqnarray}
where $a_i$ ($i=1,2,\cdots, N$) are the periods or flat coordinates for the $SU(N)$ gauge theory, satisfying $\sum_{i=1}^Na_i=0$.  The leading term scales like $\frac{1}{\epsilon_1\epsilon_2}$ and is characteristic of the saddle point behavior in the small $\epsilon_{1,2}$ limit. The Nekrasov partition function are computed by sums over Young tableaux, and in the small $\epsilon_{1,2}$ limit its logarithm is dominated by the Young tableaux that have extremal contributions. It urns out the dominant Young tableaux have the number of boxes scaling as $\frac{1}{\epsilon_1\epsilon_2}$ in the $\epsilon_{1,2}\rightarrow 0$ limit. The leading term $F^{(0,0)}$ can be computed by finding the dominant Young tableau shapes, and it was shown by this saddle point method that the leading term  $F^{(0,0)}$ is equal to the Seiberg-Witten prepotential \cite{NO}. 

It turns out the saddle point method also works when we send only one of $\epsilon$'s, say $\epsilon_2$ to zero. In this limit we consider the expansion around $\epsilon\equiv\epsilon_1$, and define the deformed prepotential $\mathcal{F}$ as
\begin{eqnarray} \label{deformedprepotential2.2}
\mathcal{F}(a_i,\epsilon) = \sum_{n=0}^{\infty}  \epsilon^{2n} F^{(n,0)}(a_i)
\end{eqnarray}
The deformed prepotential can be again computed by finding the extremal Young tableaux in the $\epsilon_2\rightarrow 0$ limit \cite{Poghossian, FMPP}. Here we will not go into the details of the derivation but simply quote the results in  \cite{Poghossian, FMPP}. For the case of $SU(N)$ theory with $N_f$ fundamental matters, the saddle point equation is 
\begin{eqnarray} \label{saddle1}
qM(x-\epsilon) w(x) w(x-\epsilon) -w(x)P(x) + 1=0 
\end{eqnarray}
The explanation of the notations follows. Here $q$ is a power of the dynamical scale for asymptotically free theories of $N_f<2N$ and the gauge coupling $q=e^{2\pi i\tau}$ for the conformal theory $N_f=2N$. The power of the $q$ parameter counts the number of instanton in the contribution to the Nekrasov partition function. The $w(x)$ is a spectral function that encodes the dominant Young tableau configuration in the $\epsilon_2\sim 0 $ limit. The $P(x)$ is a degree $N$ polynomial, and $M(x)$ is a degree $N_f$ polynomial parametrized by the mass of fundamental matters 
\begin{eqnarray}
P(x)=\prod_{i=1}^N (x-b_i),  ~~~ M(x)= \prod_{i=1}^{N_f} (x+m_i)
\end{eqnarray}
Furthermore, the parameters $b_i$ in $P(x)$ are related to the expectation value of the adjoint scalar field $\phi$ in the $\mathcal{N}=2$ gauge multiplet 
\begin{eqnarray}
\langle   \tr(\phi^J)\rangle =\sum_{i=1}^{N} b_i^J,
\end{eqnarray}
and  the deformed periods $\tilde{a}_i$ can be computed by a residue formula
\begin{eqnarray}
\tilde{a}_i = -\sum_{n=0}^{\infty} \Res_{x=b_i +n\epsilon} x\partial_x \log w(x),
\end{eqnarray}
where we use the tilde symbol to denote the period is deformed by $\epsilon$ parameter, as it turns out that it is different from the usual period $a$ in Seiberg-Witten theory. The instanton parts of the deformed prepotential is computed by a generalized Matone relation \cite{Matone}
\begin{eqnarray}
2q\frac{d\mathcal{F}_{inst}(\tilde{a}_i,\epsilon, q)}{d q} = \sum_{i=1}^N \tilde{a}_i^2 -  \langle \tr(\phi^2)\rangle =  \sum_{i=1}^N(\tilde{a}_i^2 - b_i^2).
\end{eqnarray}
We will see that at  low orders, the amplitudes $F^{(0,0)}$ and $F^{(1,0)}$ may also have some simple $q$-dependence in the classical and perturbative contributions, besides the main instanton contributions.

In this paper we consider the case of $SU(2)$ Seiberg-Witten theory whose Coulomb moduli space is described by a complex u-plane, where $u$ is the expectation value $u= \frac{1}{2}\langle \tr(\phi^2)\rangle$. It turns out that in order to ensure the SU condition $\tilde{a}_1= -\tilde{a}_2 \equiv  \tilde{a}$, we can choose the parameters $b_1= -b_2\equiv b$. So the modulus can be written $u=\frac{1}{2}(b_1^2+b_2^2)=b ^2$, and the polynomial $P(x)=x^2-u $. The residue formula and the generalized Matone relation are 
\begin{eqnarray} \label{residue2.8}
\tilde{a} = -\sum_{n=0}^{\infty} \Res_{x=b +n\epsilon} x\partial_x \log w(x), ~~~
q\frac{d\mathcal{F}_{inst}(\tilde{a}, \epsilon, q)}{d q} =  \tilde{a}^2-u
\end{eqnarray}

The authors in \cite{Poghossian, FMPP} use the saddle point equation (\ref{saddle1}) and formulae (\ref{residue2.8}) to solve the deformed prepotential $\mathcal{F}(\tilde{a},\epsilon, q)$ perturbatively in $q$ parameter and the solution is exact in $\epsilon$ parameter.  On the other hand, in order to make connection with the higher genus formulae in our paper  \cite{HK2010, HKK}, we need to instead solve the deformed prepotential exactly in $q$ parameter and but perturbatively in $\epsilon$ parameter. We will do this in the following sections.

In \cite{Chen:2011}, the authors show that the NS limit of Nekrasov function is equivalent to the F-terms of certain two-dimensional supersymmetric gauge theories, by the analysis of the corresponding saddle point equations.   
In \cite{Bonelli:2011}, the relation to the quantum Hitchin system is studied. 

In \cite{arXiv:0909.3531} the authors showed it was quite simple to prove the AGT conjecture for $SU(2)$ theory  in the Nekrasov-Shatashvili limit. This is due to the fact that the $n$-instanton contribution in the Nekrasov partition function is dominated by only one pair of Young tableau $([1^n], \emptyset)$ in this limit. In this paper we are interested in the logarithm of the Nekrasov partition function, or the free energy. It is known that the $n$-instanton contribution of the Nekrasov partition function has the leading singular behavior $\frac{1}{(\epsilon_1\epsilon_2)^n}$ in small $\epsilon_1, \epsilon_2$ limit. When one computes the logarithmic free energy, the higher order singular terms cancel out, and one finds that the  leading singular term in the free energy is only $\frac{1}{\epsilon_1\epsilon_2}$. In order to compute the leading order term of the free energy around $\epsilon_2\sim 0$, we actually need to include some sub-leading terms in the partition function.  So the Nekrasov-Shatashvili limit  of the free energy  contains more information and is much more complicated than that of the partition function.

\section{Pure $SU(2)$ theory}

As a  first step we consider the simple case of pure $SU(2)$ theory without matter. In this case the polynomial $M(x)=1$ and the saddle point equation becomes 
\begin{eqnarray} \label{saddle2}
q w(x) w(x-\epsilon) -w(x)P(x) + 1=0 
\end{eqnarray}
We write the $w(x)$ and the deformed period $\tilde{a}$ in small $\epsilon$ expansion as
\begin{eqnarray}
w(x) =\sum _{n=0}^{\infty} w_n(x) \epsilon^{n}, ~~~~ \tilde{a} = \sum_{n=0}^{\infty} a_n \epsilon^n
\end{eqnarray}
We plug the expansion of $w(x)$ into the saddle point equation (\ref{saddle2}) and solve for $w_n(x)$'s to the few orders. With $P(x)=x^2-u=x^2-b^2 $, we find 
\begin{eqnarray}
w_0(x) &=& \frac{P(x)-\sqrt{P(x)^2-4q}}{2q}, \nonumber \\
w_1(x) &=& \frac{x(P(x)-\sqrt{P(x)^2-4q})^2}{2q(P(x)^2-4q)}, \nonumber \\
w_2(x) &=& \frac{1}{2q(P(x)^2-4q)^3} ~[ P(x)^5 (P(x)-\sqrt{P(x)^2-4q})  \nonumber \\
&& -2qP(x)^2 (12x^4-16 ux^2+4u^2+(3u-11x^2)\sqrt{P(x)^2-4q})   \nonumber \\
&& +8q^2 (10x^4- 12ux^2+2u^2 +(u-4x^2)\sqrt{P(x)^2-4q})
] 
\end{eqnarray} 

At leading order $\epsilon=0$, the equation for $w_0(x)$ is a simple quadratic equation. There are two solutions for $w_0(x)$ and we choose the one with minus sign in front of the quadratic discriminant. We will also use the sign convention $P(x)>0$ when we expand the function perturbatively around $q\sim 0$. There are only rational functions of $x$ in the perturbative series expansion around $q\sim 0$, so that the residue calculations are simple to do perturbatively.  Our choice of convention for $w_0(x)$ and $P(x)>0$ gives the correct sign for the leading period $a_0=\sqrt{u} +\mathcal{O}(q)$.

Th deformed period $\tilde{a}$ can be computed perturbatively in $\epsilon$ parameter as residue around $b=\sqrt{u}$, 
\begin{eqnarray} \label{1107.3.4}
\tilde{a} = -\Res_{x=b} x\partial_x [\log(w_0(x)) + \frac{w_1(x)}{w_0(x)} \epsilon + ( \frac{w_2(x)}{w_0(x)} -\frac{w_1(x)^2}{2w_0(x)^2} )\epsilon^2 +\mathcal{O}(\epsilon^3)]
\end{eqnarray}
Here in the $\epsilon\sim 0$ limit, all possible poles at $x=b+n\epsilon$ in (\ref{residue2.8}) collapse to $x=b$, so we only need to compute the residue around $x=b$.

For an arbitrary function $f(x)$, we can compute the derivative $-xf^\prime (x) =(-xf(x))^\prime +f(x)$. If there is no branch cut for the function $f(x)$ around the residue point in the complex plane, we can ignore the total derivative and  simplify the calculations of the residue  
\begin{eqnarray}
-\Res_{x=b}  x\partial_x f(x) = \Res_{x=b} f(x) .
\end{eqnarray}
It turns out this simplification is valid for the higher order terms in (\ref{1107.3.4}) since there is no logarithmic branch cut around the residue point. But for the leading term there is a logarithmic cut $\log(w_0(x))$, so we can not use this formula.

Now we consider the leading order period $a_0 =-\Res_{x=b}  \frac{ xw_0^\prime(x)}{w_0(x)} $, which can be computed perturbatively to the first few orders around $q \sim 0$. We assume $P(x)>0$ and expand the expression for $w_0(x)$ around $q\sim 0$, and find  
\begin{eqnarray}
 a_0
&=& \Res_{x=b} [\frac{2x^2}{P(x)} +\frac{4x^2}{P(x)^3}q+\frac{12x^2}{P(x)^5}q^2 +\mathcal{O}(q^3)] \nonumber \\
&=&  \sqrt{u}(1-\frac{q}{4 u^2 }-\frac{15 q^2}{64 u^4} +\mathcal{O}(q^3) )
\end{eqnarray}
We realize the the leading order period is actually the conventional undeformed period $a\equiv a_0$ in Seiberg-Witten theory, which satisfies the Picard-Fuchs differential equation $4(4q-u^2)\partial_ u^2 a =a$.  This can be shown exactly 
\begin{eqnarray} \label{1109.3.7}
4(4q-u^2)\partial_ u^2 a - a =\Res_{x=b} \frac{d}{dx} [-\frac{2x^3(x^4 -4ux^2 +3u^2+4q)}{(P(x)^2-4q)^{\frac{3}{2}}}] =0
\end{eqnarray}
The residue vanishes since it can be written as a total derivative and there is no branch cut around the residue point $x=b$.

In general we find a contour integral or residue vanishes if the indefinite integral can be performed nicely, and the result is expressed a rational function of $x$ and the square root $\sqrt{P(x)^2-4qM(x)}$, since there is usually no branch cut in the rational functions. We have to be a little more careful if the indefinite integral involving logarithm, but this case can be easily dealt with by taking account of the branch cut of the logarithm around the contour. Otherwise, if the indefinitely integral can not be done nicely, which implies that the integral is a generic elliptic integral, there will be branch cut around the contour and the residue will not vanish. In this case we will to relate the integral to other known integrals by adding some total derivatives of rational functions of $x$ and  $\sqrt{P(x)^2-4qM(x)}$, which have no branch cut around the contour.

We compute the deformed periods to the next few orders. We find the odd terms can be always written as a total derivative with no branch cut, so the residue vanishes. For example, we find the indefinite integral 
\begin{eqnarray} \label{log3.7}
\int \frac{w_1(x)}{w_0(x)} dx = \frac{1}{4}\log(P(x)^2-4q)-\frac{1}{2} \log [P(x)+\sqrt{P(x)^2-4q} ]
\end{eqnarray}
so $-x\partial_x (\frac{w_1(x)}{w_0(x)}) =-\frac{d}{dx} [x\frac{w_1(x)}{w_0(x)}] + \frac{w_1(x)}{w_0(x)}$ is also a total derivative. Around $q\sim 0$ the leading order behavior is $\frac{w_1(x)}{w_0(x)}\sim q$, and since there is no branch cut in the logarithms in (\ref{log3.7}) for small finite $q$, the residue at  $x=b$ vanishes  $a_1=0$. Similarly we find $a_3=0$ as well because the indefinite integral can be also performed nicely.  

We compute the first non-vanishing sub-leading order contribution $a_2$ to the deformed period $\tilde{a}$. First we can compute perturbatively and find 
\begin{eqnarray} 
a_2 =  \frac{1}{\sqrt{u}} (-\frac{q}{16 u^2} -\frac{35 q^2}{128 u^4} -\frac{1155 q^3}{1024 u^6} +\mathcal{O}(q^4))
\end{eqnarray}
Then after some trials we can identify the exact formula for $a_2$ in terms of the leading undeformed period $a\equiv a_0$ as 
\begin{eqnarray}
a_2 = \frac{1}{24} (\partial_u a +2u \partial ^2_u a)
\end{eqnarray}
The exact formula can be proven by computing $a_2 - \frac{1}{24} (\partial_u a +2u \partial ^2_u a)$, and one can again show it is the residue of a total derivative of a rational function of $x$ and the square root $\sqrt{P(x)^2-4qM(x)}$, therefore vanishes. 

Similarly we compute the $\epsilon^4$ order contribution. Due to the Picard-Fuchs equation (\ref{1109.3.7}), the expression can be written in some different forms 
\begin{eqnarray}
a_4 &=&   \frac{1 }{5760} (75\partial^2_u a +120u \partial^3_u a +28u^2 \partial^4_u a)  \nonumber \\
&=& \frac{(60qu-u^3)\partial_u a +2(300q^2 +153 qu^2 -u^4)\partial^2_u a}{2880(u^2-4q)^2}
\end{eqnarray}
So we find the few order expansion for the deformed period 
\begin{eqnarray} \label{deformedperiod3.10}
\tilde{a} &=& a + a_2 \epsilon^2 +a_4 \epsilon^4 +\mathcal{O}(\epsilon^6) \\ \nonumber
&=& a+ \frac{\epsilon^2}{24}(\partial_u a+2 u\partial_u^2 a)+
\frac{\epsilon^4 }{5760} (75\partial^2_u a +120u \partial^3_u a +28u^2 \partial^4_u a) +\mathcal{O}(\epsilon^6)
\end{eqnarray}

It turns out the deformed period (\ref{deformedperiod3.10})  is the same as in the sine-Gordon quantum model studied in \cite{MM, HKK}. One can probably prove the equivalence by some ingenious changes of variables.  In \cite{MM, HKK} the deformed dual period $\tilde{a}_D = \frac{\partial \mathcal{F}(\tilde{a})}{\partial \tilde{a}}$ is used to determine the equation for the deformed prepotential. Here we will follow a different procedure and use the generalized Matone relation which has been derived from the saddle point approach in \cite{Poghossian, FMPP}. 

To simplify the analysis, we convert the derivative with respect to $q$ in the Matone relation to derivative with respect to $a$. Using dimensional analysis we see the instanton parts of the Nekrasov partition functions can be written as functions of the dimensionless combination  $\frac{q}{a^4}$ up to simple factors,  
\begin{eqnarray}
F_{inst}^{(n,0)}(a,q) = \frac{1}{a^{2n-2}}  f_n (\frac{q}{a^4}).  
\end{eqnarray}
There are also perturbative contributions 
\begin{eqnarray}
&&F^{(0,0)}_{pert}(a,q) = a^2\log(\frac{a^4}{q}), ~~~
F^{(1,0)}_{pert}(a,q) =  \frac{1}{24}\log(\frac{a^4}{q}) , 
\nonumber \\ && F^{(n,0)}_{pert}(a,q) \sim \frac{1}{a^{2n-2}}   ~~ n\geq 2.
\end{eqnarray}
Taking into account these contributions, we can write the instanton contributions in terms of the total contributions  and convert the derivatives  
\begin{eqnarray}
q\frac{d {F}_{inst}^{(0,0)}(a, q) }{d q} &=& \frac{1}{2}F^{(0,0)} -\frac{1}{4}a\frac{\partial F^{(0,0)}}{\partial a} +a^2
 \nonumber \\
q\frac{d {F}_{inst}^{(1,0)}(a, q) }{d q} &=& -\frac{1}{4} a \frac{\partial F^{(1,0)}}{\partial a} +\frac{1}{24}
 \nonumber \\
q\frac{d {F}_{inst}^{(n,0)} (a, q) }{d q} &=& \frac{1-n}{2}F^{(n,0)} -\frac{1}{4} a \frac{\partial F^{(n,0)}}{\partial a} ,~~ n\geq 2
\end{eqnarray}
We expand the generalized Matone relation by plugging the above equations in (\ref{deformedprepotential2.2}), and use  (\ref{deformedperiod3.10}) 
\begin{eqnarray} \label{deformedMatone3.11}
&& q\frac{d\mathcal{F}_{inst}(\tilde{a}, \epsilon, q)}{d q} -  \tilde{a}^2 + u   \\ \nonumber 
&=&  \frac{1}{2}F^{(0,0)}(a)  -\frac{1}{4}a\frac{\partial F^{(0,0)}(a) }{\partial a} +u 
 +\frac{\epsilon^2 }{4}
 [ a_2 (a_D+2\pi i \tau a)  - a\frac{\partial F^{(1,0)}(a) }{\partial a} + \frac{1}{6}]  \\ \nonumber 
  &&  +\frac{\epsilon^4}{2} [-F^{(2,0)}(a)-\frac{a}{2}\frac{\partial F^{(2,0)}(a) }{\partial a}  
  -\frac{a_2}{2} \partial_a (a \frac{\partial F^{(1,0)}(a) }{\partial a} ) 
  \\ \nonumber && +  \frac{a_4}{2} (a_D+2\pi i \tau a) +
   \frac{a_2^2 }{4} \partial_a (a_D+2\pi i \tau a)  ] +\mathcal{O}(\epsilon^6)
\end{eqnarray}
Here the second derivative of the prepotential is the gauge coupling $\frac{\partial^2 F^{(0,0)}(a) }{\partial^2 a} =-2\pi i\tau$, and we use the notation of the dual period $\frac{\partial F^{(0,0)}(a)}{\partial a} = a_D $. We note the definition of the parameter $\tau$ is the same as the elliptic parameter of the Seiberg-Witten curve, and is twice the convention used in \cite{HK2006, HKK}. The parameter $q$ is the 4th power of the asymptotically free scale of the pure Seiberg-Witten gauge theory, and in the followings we will no longer need to compute the derivative of $q$, so for convenience we will set $q=1$, which can always be easily recovered by dimensional analysis. The theory is then characterized by one independent parameter, the modulus parameter $u$ on the complex plane, and the other parameters  $\tau$,  $a$ and $a_D$ are functions of the modulus $u$. We will write down the functional relations between these parameters. 

The leading order equation in (\ref{deformedMatone3.11})is the conventional Matone relation. Taking derivative with respect to the period $a$ for the leading order Matone relation, we can find the formula for the dual period in terms of period $a$ and modulus $u$, 
\begin{eqnarray} \label{1109.3.15}
 a_D = -2\pi i \tau a -4\partial_a u
 \end{eqnarray}

The Seiberg-Witten curve for pure $SU(2)$ gauge theory in the Weierstrass form is $y^2 = 4x^3 -g_2(u) x- g_3(u)$, where 
\begin{eqnarray}
g_2(u) = \frac{4}{3}(u^2-3), ~~~ g_3(u) =\frac{4}{27}u(9-2u^2) 
\end{eqnarray}
The relations between the period or flat coordinate $a$ coupling $\tau$, and modulus $u$ are
\begin{eqnarray}
&& J(\tau)= \frac{E_4(\tau)^3}{E_4(\tau)^3-E_6(\tau)^2} = \frac{g_2(u)^3}{ g_2(u)^3-27g_3(u)^2} , \label{1106.3.16} \\
&& \frac{da}{du} =\sqrt{-\frac{1}{18} \frac{g_2(u)}{g_3(u)} \frac{E_6(\tau)}{E_4(\tau)}},  \label{1106.3.17}
\end{eqnarray}
see e.g. \cite{HK2010,HKK}.  In the case of pure $SU(2)$ Seiberg-Witten, we can write explicit formulae for $u$, $a$ and also $a_D$ through (\ref{1109.3.15}), as Eisenstein series and Jacobi theta functions in terms of the coupling $\tau$,
\begin{eqnarray} \label{1106.3.18}
u &=& \frac{\theta_2^4(\tau) +\theta_3^4(\tau)}{\theta_2^2(\tau) \theta_3^2(\tau) },   \nonumber \\
a &=& \frac{2 E_2(\tau) +\theta_2^4(\tau)+\theta_3^4(\tau) }{3\theta_2(\tau) \theta_3(\tau)}
\end{eqnarray}
It is straightforward to check that the formulae for $u$ and $a$ provide the solution for the relations (\ref{1106.3.16}, \ref{1106.3.17}), using the well known Ramanujan derivative identities for Eisenstein series and Jacobi moduli forms.  To compare with the formulae in the convention in \cite{HK2010, HKK}, we can use the doubling formulae for Eisenstein series and Jacobi moduli forms, and find 
\begin{eqnarray} \label{formulae3.14}
u &=& 2 \frac{\theta_3^4(\frac{\tau}{2})+ \theta_4^4(\frac{\tau}{2})}{\theta_2^4(\frac{\tau}{2})} \,,  \nonumber \\
a &=& 2 \frac{E_2(\frac{\tau}{2}) +\theta_3^4(\frac{\tau}{2})+\theta_4^4(\frac{\tau}{2}) }{3\theta_2^2(\frac{\tau}{2})} 
\end{eqnarray}
We see this is the same formulae as in  \cite{HK2010, HKK} except a factor of $2$ difference due to our convention for $u$ and $a$ here. 

The dual period is defined by $a_D=\frac{\partial F^{(0,0)}(a)}{\partial a}$, so it is also determined by the equation 
\begin{eqnarray}
\frac{d a_D}{d a}= \frac{d^2 F^{(0,0)}(a)}{d a^2} =-2\pi i \tau 
\label{1109.3.19}
\end{eqnarray}
We check two things about the dual period $a_D$, with quasi-modular formulae (\ref{1106.3.18}) and the Ramanujan derivative identities. Firstly, we can check $a_D$ satisfy the same Picard-Fuchs equation as period $a$ with respect to $u$ in (\ref{1109.3.7}). Secondly, we can verify the leading order conventional Matone relation by taking a further derivative on both sides of  (\ref{1109.3.15}) with respect to $a$, and check with (\ref{1109.3.19}).

The equations from the deformed Matone relation (\ref{deformedMatone3.11}) at order $\epsilon^2$ and $\epsilon^4$ are 
\begin{eqnarray} \label{F103.15}
a\frac{\partial F^{(1,0)}(a) }{\partial a} - \frac{1}{6} = \frac{1}{24}(\partial_u a+2 u\partial_u^2 a)  (a_D+2\pi i \tau a) 
\end{eqnarray}
\begin{eqnarray} \label{F203.16}
&& F^{(2,0)}(a)+\frac{a}{2}\frac{\partial F^{(2,0)}(a) }{\partial a}   \\  \nonumber
  &=& -\frac{a_2}{2} \partial_a (a \frac{\partial F^{(1,0)}(a) }{\partial a} ) 
  +  \frac{a_4}{2} (a_D+2 \pi i \tau a) +
   \frac{a_2^2 }{4} \partial_a (a_D+2 \pi i \tau a) 
\end{eqnarray}

In \cite{HKK} we derive higher genus formulae from holomorphic anomaly and gap conditions. The formulae for $F^{(1,0)}$ and $F^{(2,0)}$ are
\begin{eqnarray} \label{Fgformulae3.16}
F^{(1,0)}(a) &=& \frac{1}{24} \log(u^2-4) \,, \nonumber \\
F^{(2,0)}(a) &=& - \frac{u(45u X+4u^2+300)}{8640 (u^2-4)^2}  \,.
\end{eqnarray}
where $X=\frac{E_2(\tau)E_4(\tau)}{E_6(\tau)}\frac{g_3(u)}{g_2(u)}$.

Now we can check our higher genus formulae (\ref{Fgformulae3.16}) satisfy these equations (\ref{F103.15}, \ref{F203.16}) derived from the saddle point method, using formulae (\ref{1106.3.18}) and the Ramanujan derivative identities for Eisenstein series and Jacobi moduli forms. The checks are straightforward but might become tedious if done manually, so one might resort to computer algebra manipulations. Thus we have proven these higher genus formulae for $F^{(1,0)}$ and $F^{(2,0)}$.

\subsection{The deformed dual period}

The deformed period $\tilde{a}$ is the residue for contour integral (\ref{residue2.8}) of $-x\partial_x \log w(x)$ of the spectral function $w(x)$. We can define a deformed dual period $\tilde{a}_D$ as the contour integral of the same integrand but around a different B cycle
\begin{eqnarray}
\tilde{a}_D = - \frac{1}{2\pi i}\oint_B x\partial_x \log w(x)
\end{eqnarray}

   At leading order the dual period $a_{D0}\equiv a_D$ should satisfy the same Picard-Fuchs differential equation so it is the conventional dual period in Seiberg-Witten theory. The higher order contributions to the deformed dual period can be written as derivatives of leading dual period $a_D$, in the same way as the deformed period $\tilde{a}$, since the derivation of the formulae only depends on the integrand in the contour integral but not the contour. We find the same formulae as  (\ref{deformedperiod3.10}) 
\begin{eqnarray}  \label{1106.3.24}
\tilde{a}_D &=& a_D + a_{D2} \epsilon^2 +a_{D4} \epsilon^4 +\mathcal{O}(\epsilon^6) \\ \nonumber
&=& a_D+ \frac{\epsilon^2}{24}(\partial_u a_D+2 u\partial_u^2 a_D)  +
\frac{\epsilon^4 }{5760} (75\partial^2_u a_D +120u \partial^3_u a_D +28u^2 \partial^4_u a_D) +\mathcal{O}(\epsilon^6)
\end{eqnarray}

We shall show that the deformed prepotential satisfies the relation with dual deformed period 
\begin{eqnarray} \label{1106.3.25}
\frac{\partial \mathcal{F}(\tilde{a})}{\partial \tilde{a}} = \tilde{a}_D 
\end{eqnarray}
This can be probably be done with arguments similar to those of Dijkgraaf and Vafa for showing the equivalence of the prepotential of topological string theory on a Calabi-Yau manifold with a corresponding matrix model in \cite{DV2002}. Since this relation can also determine the higher order contributions of the deformed prepotential, we can prove our higher genus formulae for $F^{(n,0)}$ by showing they satisfy the relation (\ref{1106.3.25}). This is done for pure gauge theory in \cite{HKK}. Here we show the formulae again for consistency of notation and prepare for the study for the case of Seiberg-Witten theory with matters.

We expand the relation (\ref{1106.3.25}) with the formulae for deformed period   (\ref{deformedperiod3.10})  and the dual  deformed period (\ref{1106.3.24}) 
\begin{eqnarray} \label{1111.3.29}
 \frac{\partial \mathcal{F}(\tilde{a})}{\partial \tilde{a}}  - \tilde{a}_D &=&   \frac{\partial F^{(0,0)} (a)}{\partial a } - a_D  
+\epsilon^2 ( \partial_a F^{(1,0)}(a)  -2\pi i \tau a_2  -a_{D2}) \nonumber \\ &&
+\epsilon^4[ \partial_a F^{(2,0)}(a) +a_2  \partial_a^2 F^{(1,0)}(a) -2\pi i \tau a_4 -\pi i (\partial_a \tau) (a_2)^2 -a_{D4}] 
\nonumber \\ && +\mathcal{O}(\epsilon^6) 
\end{eqnarray}
The leading order is the well known Seiberg-Witten relation for the prepotential. We can again easily check that the higher genus formulae (\ref{Fgformulae3.16}) satisfy the above equations at order $\epsilon^2$ and order $\epsilon^4$, using formulae (\ref{1106.3.18}) and the Ramanujan derivative identities for Eisenstein series and Jacobi moduli forms.

\section{Seiberg-Witten theory with fundamental matters}

The expansion (\ref{1111.2.1}) of the logarithm of Nekrasov partition function has only even power terms in $\epsilon_{1,2}$. This is not actually true for the original Nekrasov function with matters. The situation can be remedied, since the odd terms can be mostly eliminated by a shift of the mass parameters of the flavor matters \cite{KW2, HKK},  
\begin{eqnarray}
m_i \rightarrow m_i  + \frac{\epsilon_1+\epsilon_2}{2}
\end{eqnarray}
Our higher genus formulae in \cite{HKK} for Seiberg-Witten theories with matters are derived based on such a shift. In the following discussion we will also make such a shift in the saddle point equation calculations to compare with the higher genus formulae. So the saddle point equation is 
\begin{eqnarray} \label{1111.4.2}
q w(x) w(x-\epsilon) \prod_{i=1}^{N_f} (x+m_i-\frac{\epsilon}{2}) -w(x)P(x) + 1=0 
\end{eqnarray}
where $P(x)=x^2-u=x^2-b^2$. 

As a main example we consider the case of one fundamental matter $N_f=1$. The calculations for the other cases $N_f=2,3,4$ to be more complicated but similar to the $N_f=1$ case.  

The calculations of $N_f=1$ case are also quite similar to those of the pure $N_f=0$ gauge theory except two technical complications. Firstly, for generic mass parameter $m_1$ we don't have close formulae for the modulus $u$ and period $a$ in terms quasi-modular forms of the elliptic parameter $\tau$ of the Seiberg-Witten curve as  in (\ref{1106.3.18}) for the $N_f=0$ case. So we have to directly deal with the functional equations (\ref{1106.3.16}, \ref{1106.3.17}). 

Secondly, because of the additional dimensional parameter $m_1$, we can not simply convert the derivative with respect to $q$ in the deformed Matone relation (\ref{deformedMatone3.11}) to the derivative with respect to $a$.  Here $q$ is the third power of the dynamical scale in the asymptotically free $N_f=1$ theory. So when we use the deformed Matone relation to compute the higher genus contributions $F^{(n,0)}$,  we need to deal with derivatives with respect to two independent variables, and the chain rule of taking derivative is more tricky in the multi-variable situation. On the other hand, as in the $N_f=0$ case, we will also use the deformed dual period to compute the higher genus contributions. In this approach there is no derivative with respect to $q$, so we can treat it as a dummy variable similarly as the mass parameter $m_1$, and we might set $q=1$ for convenience.

We write the spectral function in small $\epsilon$ expansion $w(x) =\sum _{n=0}^{\infty} w_n(x) \epsilon^{n}$ , and use the saddle point equation (\ref{1111.4.2}) to solve for $w_n(x)$'s to the few orders
\begin{eqnarray}
w_0(x) &=& \frac{P(x)-\sqrt{P(x)^2-4q(x+m_1)}}{2q(x+m_1)}, \nonumber \\
w_1(x) &=& \frac{(3x^2+4m_1x+u) [P(x)-\sqrt{P(x)^2-4q(x+m_1)}]^2}{8q(x+m_1)(P(x)^2-4q(x+m_1))}, \nonumber \\
& \cdots & \nonumber
\end{eqnarray} 

The deformed period is computed $\tilde{a}= \sum _{n=0}^{\infty} a_n \epsilon^{n}  = -\Res_{x=b} x\partial_x \log w(x)  $. The leading order period $a\equiv a_0$ can be computed perturbatively for small $q$, 
\begin{eqnarray} 
a =  \sqrt{u} - \frac{m_1}{4u^{\frac{3}{2}}} q+\frac{3(u-5m_1^2)}{64u^{\frac{7}{2}}} q^2 +\frac{35m_1(u-3m_1^2)}
{256u^{\frac{11}{2}}}q^3 +\mathcal{O}(q^4)
\end{eqnarray}
The Picard-Fuchs equation was known in \cite{Ohta:1996} some time ago
\begin{eqnarray} \label{1111.PF}
&&  2\Delta (u) (4m_1^2-u) \partial_u^3 a +2[2\Delta(u) +(4m_1^2-3u)\partial_u \Delta(u) ]\partial_u^2 a \nonumber \\
&& +4(6u^2-18m_1^2u +8m_1^4+9m_1q)\partial _u a = 0, 
\end{eqnarray}
where $\Delta(u) $ is the discriminant 
\begin{eqnarray}
\Delta(u)  = -16u^3 +16m_1^2 u^2+72 m_1 q u -64 m_1^3q-27 q^2
 \end{eqnarray}
We check the Picard-Fuchs equation exactly by showing the left hand side is the residue of a total derivative without branch cut around the residue point, and therefore vanish. 

We find the odd terms can be written as the residue of a total derivative without branch cut around the residue point, and therefore vanish. We check this for $a_1$ and $a_3$ contributions. For example, we find the indefinite integral
\begin{eqnarray}
\int \frac{w_1(x)}{w_0(x)} dx = \frac{1}{4}\log[P(x)^2-4q(x+m_1)] - \frac{1}{2}\log[\sqrt{P(x)^2-4q(x+m_1)} + P(x)]
\end{eqnarray}
There are actually branch cut contributions in the logarithmic functions around the residue point $x=b=\sqrt{u}$ if $q=0$, but they cancel out. We find 
\begin{eqnarray}
a_1 =-\Res_{x=b} x\partial_x  \frac{w_1(x)}{w_0(x)} = \Res_{x=b} \frac{w_1(x)}{w_0(x)} =0
\end{eqnarray}
Similarly we find $a_3=0$. 

We identify the exact formulae for the non-vanishing sub-leading even terms in the expansion of the deformed period $\tilde{a}= \sum _{n=0}^{\infty} a_n \epsilon^{n}$,  
\begin{eqnarray} \label{1111.a2}
a_2 &=& \frac{(3u-2m_1^2)\partial_u a +(6u^2 -4m_1^2 u - 9m_1 q) \partial_u^2 a }{12(3u - 4m_1^2)}, \\
a_4 &=& \frac{1}{(4m_1^2 - 3u)\Delta(u)^2} \{ [  288 u^6   +480 m_1^2 u^5 +48 m_1(4 m_1^3-159 q)   u^4  \nonumber \\ && 
+(8880 m_1^3 q-64 m_1^6+7290 q^2) u^3  -144 m_1^2 q (89 m_1^3+45
   q)   u^2 \nonumber \\ &&  +6 m_1 q (4254 m_1^3 q+640 m_1^6-81 q^2) u-15 m_1^3
   q^2 (320 m_1^3+1269 q) ]\frac{\partial_u a}{180}  \nonumber \\ &&
   +   [  +4608 u^7 +7680 m_1^2 u^6  +768  (4 m_1^4-267 m_1 q) u^5 \nonumber \\ &&
    + 16 (21840 m_1^3 q-64 m_1^6+24705 q^2) u^4   -1008 m_1^2 q (464 m_1^3+933
   q)   u^3    \nonumber \\ &&  +72 m_1 q  (28232 m_1^3 q+2176 m_1^6-1323
   q^2) u^2  -3 q^2  (313200 m_1^3 q+361984 m_1^6   \nonumber \\ &&  +149445 q^2) u +192 m_1^2 q^2 (-1341 m_1^3
   q+1600 m_1^6+6804 q^2) ]\frac{\partial_u^2 a}{1440}  \}  \label{1111.a4}
\end{eqnarray} 
It takes some trials to identify the formulae. Due to the Picard-Fuchs equation (\ref{1111.PF}), it is sufficient to write the higher order period as a linear combination of $\partial_u a$ and $\partial_u^2 a$, and there is no need for higher derivatives. The formulae can be again easily proven by subtracting the two sides of the equations and showing that the result is a contour integral of a total derivative of a rational function of $x$ and the square root $\sqrt{P(x)^2-4qM(x)}$,  without branch cut around the contour. 

The Seiberg-Witten curve in elliptic form is $y^2=4x^3-g_2(u) x-g_3(u)$, where 
\begin{eqnarray}
g_2(u) =\frac{4u^2}{3}-4m_1q,~~~ g_3(u) = \frac{8u^3}{27} +\frac{4}{3}m_1 qu-q^2 
\end{eqnarray}
To write the higher genus formulae, we introduce the elliptic parameter $\tau$ of the curve, which is also related to the prepotential as $\partial_a^2 F^{(0,0)}(a) = -2\pi i \tau$.  The relations between the period  $a$,  coupling $\tau$, and modulus $u$ are captured by the functional equations similar to the pure gauge theory,  
\begin{eqnarray}
&& J(\tau)= \frac{E_4(\tau)^3}{E_4(\tau)^3-E_6(\tau)^2} = \frac{g_2(u)^3}{ g_2(u)^3-27g_3(u)^2} , \label{1111.4.11a} \\
&& \frac{da}{du} =\sqrt{-\frac{1}{18} \frac{g_2(u)}{g_3(u)} \frac{E_6(\tau)}{E_4(\tau)}}, \label{1111.4.12a}
\end{eqnarray}
The Picard-Fuchs equation (\ref{1111.PF}) between $a$ and $u$ can be derived from these functional relations. 

The formula (\ref{1109.3.15}) for the dual period $a_D =\partial_a F^{(0,0)}(a)$ in the pure gauge theory case is no longer valid for the case of theories with matters here. The functional relation of $a_D$ with the other parameters is determined by $\partial _a a_D =\partial_a^2 F^{(0,0)}(a) =-2\pi i\tau$. In terms of the modulus $u$ we can write 
\begin{eqnarray} \label{1111.aD}
 \frac{da_D}{du} = -2\pi i \tau \frac{d a}{d u} = -2\pi i \tau \sqrt{-\frac{1}{18} \frac{g_2(u)}{g_3(u)} \frac{E_6(\tau)}{E_4(\tau)}}
 \end{eqnarray}
The dual period $a_D$ satisfies the same Picard-Fuchs differential equation with respect to $u$ as the period $a$.

In \cite{HKK} we derive the higher genus formulae from holomorphic anomaly equations and boundary gap conditions. For example, we found 
\begin{eqnarray} 
F^{(1,0)} &=& \frac{1}{24}\log(\Delta(u)/q^2)   \label{1111.4.14}  \\
F^{(2,0)} & =& \frac{1}{540\Delta(u)^2} \{ -45( 6u^2 -4m_1^2 u-9m_1q)^2 X +72 u^5 +624 m_1^2 u^4  \nonumber \\ &&
- (6372 m_1 q + 64 m_1^4)  u^3   +216 q  (28 m_1^3+45
   q) u^2 \nonumber \\ && -12 m_1^2 q  (400 m_1^3+567 q) u
+ 54 m_1 q^2 (184 m_1^3-189 q)\}  \label{1111.4.15} 
\end{eqnarray}
where $X=\frac{E_2(\tau)E_4(\tau)}{E_6(\tau)} \frac{g_3(u)}{g_2(u)}$.

As we mentioned, in the saddle point method studied here, the higher genus contributions can be calculated by two different ways: use the deformed dual period or use the deformed Matone relation. We consider these two approaches respectively to prove our higher genus formulae (\ref{1111.4.14}, \ref{1111.4.15} ).

\subsection{Use the deformed dual period}

This approach works similarly as the pure gauge theory case. We see that the sub-leading order contributions to the deformed period and the dual can be computed by the same formulae
\begin{eqnarray}
\tilde{a} &=& [1+\epsilon^2 L_2(u)  +\epsilon^4 L_4(u)+\mathcal{O}(\epsilon^6) ]a, \nonumber \\
\tilde{a}_D &=&  a_D+\epsilon^2 a_{D2}  +\epsilon^4 a_{D4} +\mathcal{O}(\epsilon^6) \nonumber \\
&=& [1+\epsilon^2 L_2(u)  +\epsilon^4 L_4(u)+\mathcal{O}(\epsilon^6) ]a_D
\end{eqnarray}
where $L_2(u)$ and $L_4(u)$ are some differential operators involve derivatives with respect to $u$, and can be found in the formulae (\ref{1111.a2}, \ref{1111.a4}). 

We can compute the higher genus contributions $F^{(n,0)}(a)$ in the deformed prepotential $\mathcal{F}(a,\epsilon) = \sum _{n=0}^{\infty} F^{(n,0)}(a) \epsilon^{2n} $ by the relation 
\begin{eqnarray}
\frac{\partial  \mathcal{F}(\tilde{a},\epsilon)}{\partial \tilde{a}} =\tilde{a}_D
\end{eqnarray}
Similar to the pure gauge theory as in (\ref{1111.3.29}), we expand the equation for small  $\epsilon$, and find the order $\epsilon^2$ and order $\epsilon^4$ equations 
\begin{eqnarray}  \label{1118.dual}
 \partial_a F^{(1,0)}(a)  &=& 2\pi i \tau a_2  + a_{D2}  \nonumber \\ 
 \partial_a F^{(2,0)}(a) &=&  -a_2  \partial_a^2 F^{(1,0)}(a) + 2\pi i \tau a_4 + \pi i (\partial_a \tau) (a_2)^2 + a_{D4}
 \end{eqnarray}
Using the functional relations (\ref{1111.4.11a} , \ref{1111.4.12a}, \ref{1111.aD}) and Ramanujan derivative identities,   we check our higher genus formulae (\ref{1111.4.14}, \ref{1111.4.15} ) satisfy these equations.

\subsection{Use the deformed Matone relation}

The deformed Matone relation is 
\begin{eqnarray} \label{1111.deformedMatone}
q\frac{d\mathcal{F}_{inst} (\tilde{a}, \epsilon, q)}{d q} -  \tilde{a}^2 + u=0
\end{eqnarray} 
Here in the equation we write only the instanton contribution to the prepotential. On the other hand, our higher genus formulae in (\ref{1111.4.14}, \ref{1111.4.15} )  include both the perturbative and instanton contributions. For $n\geq 2$, the perturbative part of the higher genus contributions $F^{(n,0)}$ is independent of the parameter $q\equiv \Lambda^{4-N_f}$, where $\Lambda$ is the asymptotically free scale of $SU(2)$ Seiberg-Witten theory with $N_f$ flavors, so it doesn't affect the deformed Matone relation whether we use the total or instanton contributions. However, for the low order $F^{(0,0)}$ and $F^{(1,0)}$, the perturbative contributions have $q$ dependence due to the logarithmic functions and we need to take in account their contributions. Specifically, the perturbative contributions are 
\begin{eqnarray} 
F^{(0,0)}_{pert}  (a,q)&=& - \frac{3}{2}(4-N_f)a^2 +\frac{3}{2}\sum_{i=1}^{N_f}m_i^2 +2a^2 \log(-4a^2/\Lambda^2) 
\nonumber \\ &&
-\frac{1}{2}\sum_{i=1}^{N_f} \{(a-m_i)^2\log[(-a+m_i)/\Lambda]  + (a+m_i)^2\log[(a+m_i)/\Lambda] \},
\nonumber \\
F^{(1,0)}_{pert} (a,q) &=& \frac{1}{12} \log(-4a^2/\Lambda^2) + \sum_{i=1}^{N_f} \frac{1}{24} \log[(-a^2+m_i^2)/\Lambda^2],
\end{eqnarray}   
where we add powers of $\Lambda$ to cancel the mass dimension of the logarithm. We compute the perturbative contributions 
\begin{eqnarray} \label{1111.pert}
&& q\frac{d F^{(0,0)}_{pert} (a,q) }{dq} =\frac{1}{4-N_f} \Lambda\frac{d F^{(0,0)}_{pert}  (a,q) }{d\Lambda} = -a^2 
 +   \frac{1}{4-N_f} \sum_{i=1}^{N_f}m_i^2 
\nonumber \\
&& q\frac{d F^{(1,0)}_{pert}  (a,q) }{dq} = \frac{1}{4-N_f}  \Lambda\frac{d F^{(1,0)}_{pert}  (a,q)}{d\Lambda} 
= -\frac{N_f+2}{12(4-N_f)}
\end{eqnarray}

We expand the deformed Matone relation (\ref{1111.deformedMatone}) to the first few orders, taking account of the perturbative contributions (\ref{1111.pert}), and use (\ref{1111.a2}, \ref{1111.a4}). We find 
\begin{eqnarray} \label{1111.4.22.deformedMatone}
&& q\frac{d\mathcal{F}_{inst} (\tilde{a}, \epsilon, q)}{d q} -  \tilde{a}^2 + u   \\
&=& [ q\partial_q F^{(0,0)} (a,q) - \frac{1}{4-N_f} \sum_{i=1}^{N_f}m_i^2  +u]   +\epsilon^2 [ q\partial_q F^{(1,0)}(a,q) 
+ \frac{N_f+2}{12(4-N_f)}+a_2 q\partial_q a_D] \nonumber \\
&& + \epsilon^4 q [\partial_q F^{(2,0)}(a,q) +a_2 \partial_a\partial_q F^{(1,0)}(a,q) +a_4 \partial_q a_D - a_2^2 \partial_q (\pi i \tau) ] +\mathcal{O}(\epsilon^6) \nonumber
\end{eqnarray}
where we have used $a_D=\partial_a F^{(0,0)}(a,q)$. 

In the conformal case $N_f=4$ the parameter $q=e^{2\pi i \tau_0} $ is related to the bare gauge coupling $\tau_0$ and is dimensionless . We note here that the bare coupling $\tau_0$ is renormalized by instanton contributions, and is different from the $\tau$ of the Seiberg-Witten curve in the functional relations (\ref{1111.4.11a} , \ref{1111.4.12a}).  We see that  if we naively set $N_f=4$ in the expressions (\ref{1111.pert}) and  (\ref{1111.4.22.deformedMatone}), they become  singular. 
This is a hint that as it turns out, the deformed Matone relation is slightly modified in the $N_f=4$ theory compared to the  asymptotically free theories. Here we consider the deformed Matone relation (\ref{1111.4.22.deformedMatone}) for the asymptotically free cases $N_f\leq 3$, and leave the discussion of the $N_f=4$ case to next section.

Our higher genus formulae (\ref{1111.4.14}, \ref{1111.4.15} )  for the $N_f=1$ case are expressed in terms of Eisenstein series $E_n(\tau)$ and modulus $u$. The partial derivatives of the these variables and also the dual period $a_D$ with respect to period $a$ can be found from the functional relations (\ref{1111.4.11a} , \ref{1111.4.12a}, \ref{1111.aD}). In order to check the higher genus formulae satisfy  the deformed Matone relation (\ref{1111.4.22.deformedMatone}), we must also compute the partial derivatives with respect to $q$ parameter. To do this we first assume the validity of the leading order equation in (\ref{1111.4.22.deformedMatone}), and take the partial derivative with respect to $a$ once and twice. We find 
\begin{eqnarray}  \label{1111.4.23}
\partial_q a_D(a,q)  &=& -\frac{1}{q}\partial _a u =  - \frac{2}{q} \frac{(u^2-3m_1q)^{\frac{1}{4}} }{E_4(\tau)^{\frac{1}{4}} }  \\
\partial_q \tau (a,q)  &=& \frac{1}{ 2\pi i q}\partial ^2_a u  =  \frac{8(u^2-3m_1q)}{3q\Delta(u)E_4(\tau)} \{
(8m_1^2-6u) E_2(\tau) \nonumber \\ && 
+ (9m_1q +4m_1^2u-6u^2 ) \sqrt{\frac{E_4(\tau) } {u^2-3m_1q }}  \} \label{1111.4.24}
\end{eqnarray}
The partial derivative of $u$ can be found by taking derivative with respect to $q$ on both sides of (\ref{1111.4.11a}) and use (\ref{1111.4.24}) for $\partial_q\tau$. We find 
\begin{eqnarray}
\partial_q u(a,q) = \frac{u}{3q} -\frac{E_2(\tau)}{3q} \sqrt{\frac{u^2-3m_1q }{E_4(\tau) }}  \label{1111.4.25}
\end{eqnarray}  
 
In deriving the formulae (\ref{1111.4.23}, \ref{1111.4.24}, \ref{1111.4.25}) we use the leading order  Matone relation  $ q\partial_q F^{(0,0)} (a,q) -\frac{m_1^2}{3}+u=0 $ in (\ref{1111.4.22.deformedMatone}). We can turn around and make a check on this equation by computing $\frac{\partial^2 u(a,q)}{\partial a\partial q }$ in two ways. Firstly,  we can compute the derivative of $q$ first using (\ref{1111.4.25}), then compute the derivative of $a$ using  (\ref{1111.4.11a} , \ref{1111.4.12a}). Secondly, we can compute the derivative of $a$ first using  (\ref{1111.4.12a}), then compute the derivative of $q$ using ( \ref{1111.4.24}, \ref{1111.4.25}). We find the same result and  therefore confirm the validity of leading order equation in ((\ref{1111.4.22.deformedMatone}). 

We can now check the order $\epsilon^2$ and $\epsilon^4$ equations in (\ref{1111.4.22.deformedMatone}). We first use the formulae (\ref{1111.4.23}, \ref{1111.4.24}, \ref{1111.4.25}) to compute the  derivatives with respect to $q$,  then we compute the derivatives with respect to $a$ using the formulae (\ref{1111.4.11a} , \ref{1111.4.12a}).  We confirm our higher genus formulae (\ref{1111.4.14}, \ref{1111.4.15} ) satisfy the  $\epsilon^2$ and $\epsilon^4$ equations in (\ref{1111.4.22.deformedMatone}).  Therefore we prove these higher genus formulae.

\subsection{The results for $N_f=2, 3,4 $}

The studies of $N_f=2,3,4$ cases are similar to the $N_f=1$ case with the addition of two technical points. Firstly, it turns out that the polynomial $P(x)$ in the saddle point equation  (\ref{1111.4.2}) is no longer simply $P(x)=x^2 - u$. Instead, the parameters are shifted by the flavor mass in order to match the convention of Nekrasov function, which is also used in our previous papers \cite{HK2009,HKK}. The correct expressions for $P(x)$ can be derived from the well known Seiberg-Witten curves \cite{SW1, SW2}. Secondly, as we mentioned, the deformed Matone relation (\ref{1111.4.22.deformedMatone})  is slightly modified for the $N_f=4$ case. 

The $SU(2)$ Seiberg-Witten curves can be written in either quartic form or the Weierstrass form. In the quartic form, the curves are $y^2=P(x)^2 -4q M(x)$, where $q=\Lambda^{4-N_f}$ for $N_f\leq 3$ and $q=e^{2\pi i\tau_0}$ for $N_f=4$ with  $\tau_0$ the bare UV gauge coupling. 
The expressions for $P(x)$ in various cases are 
\begin{eqnarray} \label{1118.4.26}
 N_f=0,1: && P(x)=x^2 - u, \nonumber \\
 N_f=2:~~  && P(x) =x^2-u +\frac{q}{2}, \nonumber \\
 N_f=3:  ~~&& P(x)=x^2 -u + q(x+\frac{p_1}{2}),  \nonumber \\
 N_f=4:  ~~ && P(x)= (1 + q)x^2 -u + q p_1 x - \frac{1}{2} p_1^2 + (1 + \frac{q}{2})p_2 ,
\end{eqnarray}
and we have used the symmetric polynomials $p_k$ for mass parameters in $M(x)$ defined as 
\begin{eqnarray}
M(x)= \prod_{i=1}^{N_f}(x+m_i)\equiv \sum_{k=0}^{N_f} p_k x^{N_f-k}. 
\end{eqnarray}

We can transform any quartic curve $y^2= c_0x^4 +4c_1x^3 +6 c_2 x^2 +4 c_3 x+ c_4$ into the Weierstrass form 
\begin{eqnarray}
y^2 = 4x^3 -g_2 x-g_3
\end{eqnarray}
where the expressions for $g_2$ and $g_3$ are 
\begin{eqnarray} \label{1118.4.29}
g_2 &=& c_0c_4 - 4c_1c_3+3c_2^2 ,  \nonumber \\
g_3 &=& c_0 c_2 c_4+2c_1 c_2 c_3 - c_0c_3^2 - c_1^2 c_4 - c_2^3
\end{eqnarray}
The $g_2$ and $g_3$ functions in the Weierstrass form of the Seiberg-Witten curve are used in \cite{HK2009, HKK} to provide the functional relations (\ref{1111.4.11a} , \ref{1111.4.12a}) between various parameters and to compute the higher genus formulae. When we transform the quartic curves $y^2=P(x)^2 -4q M(x)$ with $P(x)$ given by (\ref{1118.4.26}) to the  Weierstrass form according to (\ref{1118.4.29}), we find they are the same curves used in \cite{HK2009, HKK}. The curve for $N_f=4$ is also the same as the hyper-elliptic curve in \cite{Argyres} for $SU(N)$ with $N_f=2N$ specializing to $SU(2)$, after a redefinition of the $u$ parameter. 

At the leading order $\epsilon=0$,  the saddle point equation (\ref{1111.4.2}) is a simple quadratic equation for the spectral function $w(x)$. The discriminant $y=\sqrt{P(x)^2 -4q M(x)}$ of the quadratic equation  is the same as the Seiberg-Witten curve in quartic form if we identify the $P(x)$ of the saddle point equation  (\ref{1111.4.2}) with that of the Seiberg-Witten curve. Therefore we see the formulae (\ref{1118.4.26}) for $P(x)$  in the quartic Seiberg-Witten curves provide the correct expressions for the polynomial $P(x)$ in the saddle point equation (\ref{1111.4.2}) in various cases.  

We compute the leading order period $a$ using the residue formula 
\begin{eqnarray} \label{1119.4.30}
a=-\Res_{x=b} x\partial_x \log(w_0(x))
\end{eqnarray}
Here we note the residue point $x=b$ is a root of $P(x)=0$ for (\ref{1118.4.26}) and for $N_f\geq 2$ cases it is not simply $\sqrt{u}$ but a little more complicated.  We expand perturbatively around small $q\sim 0$ and check the asymptotic expansion of the period $a$ is the same as that calculated from the relations  (\ref{1111.4.11a} , \ref{1111.4.12a}). It should be straightforward to prove the equivalence exactly, by deriving a Picard-Fuchs differential equation for $a$ in terms of $u$ from the elliptic integral and show it is the same equation implied by the relations (\ref{1111.4.11a} , \ref{1111.4.12a}). The Picard-Fuchs equation for the $N_f=2,3$ cases were studied some time ago in \cite{Ohta2}.

Similar to the $N_f=1$ case we find the odd terms $a_1, a_3$ vanish for $N_f=2,3,4$, because the corresponding indefinite integrals can be performed nicely in terms of simple functions. 

We compute the non-vanishing sub-leading contributions $a_2, a_4$, and find the formulae in terms of some differential operators acting on the leading order period $a$. it is too complicated to write down all the formulae. Here as a sample, we provide formulae for $a_2$ in the simple case when only one of the hypermultiplets has non-zero mass $m$,
\begin{eqnarray}
N_f=2:   && a_2 =  \frac{1}{12 (4 u^2 - 6 m^2 u -q^2)} \{ 2(4u^2  - 3m^2 u -q^2) \partial_u a \nonumber \\ && ~~
+ (16 u^3 - 12 m^2 u^2   - 4 q^2 u  +9 m^2 q^2   )\partial_u^2 a \}, \\
N_f=3:   && a_2 =  \frac{1}{24(u-2 m^2) } \{ 2(5u  - 4m^2 ) \partial_u a \nonumber \\ && ~~
+ (20 u^2- (16 m^2 + q^2) u  - m^2 q^2   )\partial_u^2 a \}, \\
N_f=4:   && a_2 =  - \frac{1}{24 m^2 } \{ 2(6u  - 2m^2  +q m^2) \partial_u a \nonumber \\ && ~~
+ (24 u^2 +2 m^2 (2+5 q) u  +m^4 (-4+5 q+q^2)   )\partial_u^2 a \}.  \label{1118.4.33}
\end{eqnarray}
Again the deformed dual period $\tilde{a}_D$ are computed by the same formulae as the deformed period $\tilde{a}=\sum_{n=0}^{\infty} a_n\epsilon^n $.

We note that there is no problem taking the massless limit $m\rightarrow 0$ in the asymptotically free $N_f\leq 3$ cases. However, for the $N_f=4$ case, we see that the formula (\ref{1118.4.33}) for $a_2$ is singular in the massless limit $m\rightarrow 0$. We encounter the same phenomenon in \cite{HKK}. The discriminant behaves like $\Delta\sim u^6$ in the $N_f=4$ massless limit. All 6 discriminant points in the $u$-plane collide at $u=0$, and some mutually non-local charged particles become massless at this point. We find the gap conditions break down, and we have to solve the higher genus amplitudes by deforming away from the massless limit. Once we have the higher genus formulae in the massive region, we can then take the massless limit for the  formulae which turns out to be non-singular. However since the formulae for the deformed period, e.g. $a_2$ in (\ref{1118.4.33}),  is singular in the massless limit, we can not directly test our massless higher genus formulae for the $N_f=4$ theory using the saddle point method.

The formulae for $F^{(1,0)}$ is simple to write 
\begin{eqnarray}
F^{(1,0)} =\frac{1}{24}\log(\Delta) - \frac{N_f+2}{12(4-N_f) }\log q
\end{eqnarray} 
where the discriminant of the Seiberg-Witten curve is $\Delta =g_2^3-27g_3^2$. We have  added a term of $\log (q)$ to cancel the mass dimension of $\Delta$ inside the logarithm for the $N_f\leq 3$ asymptotically free theories and it does not affects the equations (\ref{1118.dual}) for deformed dual period.  This $\log (q)$  term will cancel the constant $\frac{N_f+2}{12(4-N_f) }$ in the order $\epsilon^2$ equation in the deformed Matone relation (\ref{1111.4.22.deformedMatone}). The formulae for $F^{(2,0)}$ in the $N_f=2,3,4$ cases are similar to that of the $N_f=1$ case (\ref{1111.4.15}) but more complicated to write down here.

We check the higher genus formulae for $F^{(1,0)}$ and $F^{(2,0)}$ for $N_f=2,3,4$ theories satisfy the order $\epsilon^2$ and $\epsilon^4$ equations (\ref{1118.dual}) from the deformed dual period $\tilde{a}_D$. For the asymptotically free $N_f=2,3$ theories, we also check the higher genus formulae satisfy the order $\epsilon^2$ and $\epsilon^4$ equations in the deformed Matone relation   (\ref{1111.4.22.deformedMatone}).

Now we discuss the deformed Matone relation for the $N_f=4$ theory. The gauge coupling is renormalized and its formula for the massless case in terms of modular forms is given in \cite{Grimm:2007}. First we consider the leading order equation, and we find the equation is modified as the following  
\begin{eqnarray} \label{1119.4.35}
q\partial_q F^{(0,0)}_{inst} (a,q) -a^2+\frac{p_1^2 -(2-q)p_2}{2(1-q)} +\frac{u}{1-q} =0
\end{eqnarray}
where we see the $u$ term is modified by a factor of $(1-q)$ compared to that of the asymptotically free theories (\ref{1111.4.22.deformedMatone}), and there is also a mass term. The modulus $u$ can be expressed as perturbative series of $q$ in terms of period $a$, by the relations  (\ref{1111.4.11a} , \ref{1111.4.12a}) or by inverting the formula (\ref{1119.4.30}) for the period $a_0\equiv a$.  We check the leading order Matone equation 
(\ref{1119.4.35}) perturbatively using the Nekrasov  instanton formula for the prepotential $F^{(0,0)}_{inst} (a,q)$ and the perturbative expansion for $u(a,q)$.  

Since for the $N_f=4$ theory, the parameter $q$ is dimensionless, we can not simply recover $q$ dependence in the perturbative contributions  by dimensional analysis. Here we will assume the $q$-dependence in the perturbative contributions only comes from $F^{(0,0)}$ and $F^{(1,0)}$ and is rather trivial. 

In addition to the perturbative and instanton contributions, there are also $q$-dependence in the classical contributions in the $N_f=4$ theory, 
\begin{eqnarray}
F_{classical}^{(0,0)} (a,q) =-2\pi i \tau_0 a^2 =-\log(q) a^2 
\end{eqnarray} 
The classical contribution accounts for the $-a^2$ term in (\ref{1119.4.35}). So the total contribution for the prepotential $F^{(0,0)}$ satisfy the Matone relation 
\begin{eqnarray} \label{1119.leading}
 q\partial_q F^{(0,0)}(a,q)  +\frac{p_1^2 -(2-q)p_2}{2(1-q)} +\frac{u}{1-q} =0
\end{eqnarray}

The proposal for the deformed Matone relation is to replace the period $a$ in the conventional Matone relation 
(\ref{1119.leading}) with the deformed period $\tilde{a}$, and the prepotential $F^{(0,0)}(a,q)$ with the deformed prepotential $\mathcal{F}(\tilde{a},q,\epsilon) = \sum_{n=0}^{\infty} F^{(n,0)}(\tilde{a},q) \epsilon^{2n}$. We expand the deformed Matone relation for small $\epsilon$ to find differential equations for the higher genus amplitudes 
\begin{eqnarray} \label{1119.deformedNf4}
 && q\partial_q \mathcal{F}(\tilde{a},q,\epsilon)  +\frac{p_1^2 -(2-q)p_2}{2(1-q)} +\frac{u}{1-q}    \\
  && = [q\partial_q F^{(0,0)}(a,q)  +\frac{p_1^2 -(2-q)p_2}{2(1-q)} +\frac{u}{1-q}] 
    +\epsilon^2 q[ \partial_q F^{(1,0)}(a,q) +a_2 \partial_q a_D] \nonumber \\
&& + \epsilon^4 q [\partial_q F^{(2,0)}(a,q) +a_2 \partial_a\partial_q F^{(1,0)}(a,q) +a_4 \partial_q a_D - a_2^2 \partial_q (\pi i \tau) ] +\mathcal{O}(\epsilon^6) \nonumber
\end{eqnarray}
We find that except for a constant at order $\epsilon^2$ equation, the higher order equations are exactly the same as those of the asymptotically free case (\ref{1111.4.22.deformedMatone}), since the modified $\frac{u}{1-q}$ term does not contribute in higher orders. 

We can use the leading order Matone relation (\ref{1119.leading}) to calculate the partial derivatives of $a_D, \tau, u$ parameters with respect to $q$, similarly as in the asymptotically free theories. Taking partial derivative with respect to $a$ once and twice on both sides of (\ref{1119.leading}), we find 
\begin{eqnarray}
\partial_q a_D (a,q) = - \frac{\partial_a u}{q(1-q)}, ~~~  \partial_q \tau(a,q)  =  \frac{\partial_a^2 u}{2\pi iq(1-q)}, \label{1119.4.39}
\end{eqnarray}
which differ from the the asymptotically free theories (\ref{1111.4.23}, \ref{1111.4.24}) only by a factor of $(1-q)$. As in the asymptotically free theories, the partial derivative of $\partial_q u(a,q)$ can be found by taking derivative with respect to 
$q$ on both sides of (\ref{1111.4.11a}) and use (\ref{1119.4.39}) for $\partial_q\tau$.

We check the higher genus formulae for $F^{(1,0)}$ and $F^{(2,0)}$ with the order $\epsilon^2$ and $\epsilon^4$ equations in the deformed Matone relation (\ref{1119.deformedNf4}). For $F^{(1,0)}$ we use the formula 
\begin{eqnarray}
F^{(1,0)} =\frac{1}{24}\log(\Delta) -\frac{1}{12}\log [q(1 - q)^4]
\end{eqnarray} 
We note that the $ -\frac{1}{12} \log q(1 - q)^4$ term does not exactly match the Nekrasov instanton partition function. In order to match Nekrasov's formula, we would need to use the term $ -\frac{1}{12} \log q (1 - q)^{-2}$. The slight difference of $\frac{1}{2}\log (1-q)$ can probably be explained by some perturbative contributions to $F^{(1,0)}$.  Other than this rather trivial term, we confirm our higher genus formulae for $F^{(1,0)}$ and $F^{(2,0)}$ for the $N_f=4$ theory satisfy the deformed Matone relation.

\section{Seiberg-Witten theory with an adjoint matter}

We consider the $SU(2)$ Seiberg-Witten theory with an adjoint hypermultiplet, known as the $\mathcal{N}=2^*$  theory. The theory has $\mathcal{N}=4$ supersymmetry and is exactly conformal invariant if the adjoint hypermultiplet is massless.  

The saddle point equation in \cite{FMPP} looks quite different from the cases with fundamental flavors. One difficulty is that the mass parameter of the adjoint multiplet appears in the argument of the spectral function $w(x)$ in the saddle point equation, making it difficult to to solve the spectral function even in the leading order $\epsilon=0$. Instead, we propose to use an alternative saddle point equation similar to those of Seiberg-Witten theory with fundamental matters, by reverse engineering from the Weierstrass form of the Seiberg-Witten curve that was used in \cite{HKK} to solve the the  higher genus amplitudes. 

We find the quartic curve whose transformation according to (\ref{1118.4.29}) gives rise to the Weierstrass curve  for the $\mathcal{N}=2^*$  theory used in  \cite{HKK}.  There are actually 3 solutions up to the translation for the $x$ parameter. We use the simplest quartic curve $y^2=P(x)^2-4 q M(x)$, where 
\begin{eqnarray} \label{1122.adjointquartic}
P(x) = (1 + q)x^2 -\frac{1 + q}{2} m^2 - u, ~~ M(x) =x^2(x + \frac{m}{2})(x - \frac{m}{2}).
\end{eqnarray}
Here $m$ the mass of the adjoint hypermultiplet, and $q$ is related to the bare UV coupling $\tau_0$ of the theory by Jacobi theta functions 
\begin{eqnarray} \label{1122.5.2}
q=\frac{\theta_2^4(\tau_0)}{\theta_3^4 (\tau_0)} 
\end{eqnarray} 
We note this is different from the $N_f=4$ theory where the $q$ parameter in Seiberg-Witten curve is simply the exponential of the bare coupling $\tau_0$. We also use the notation $q_0=e^{2\pi i \tau_0}$ for the  $\mathcal{N}=2^*$  theory, and we can use either $q_0$ or $\tau_0$ as the argument in theta functions $\theta_i (q_0) \equiv \theta _i(\tau_0)$ ($i=2,3,4$). 

The transformation of the quartic curve to Weierstrass curve  according to (\ref{1118.4.29}) is  $y^2=4x^4 -g_2 x-g_3$ where 
\begin{eqnarray}
g_2 &=& \frac{1}{12} \{m^4 (1 - q^2 + q^4) + 4 m^2 (2 - q - q^2 + 2 q^3) u + 16 (1 - q + q^2) u^2\}, \nonumber \\
g_3 &=& - \frac{1}{432} \{ m^6 (2 - 3 q^2 - 3 q^4 + 2 q^6) +
    12 m^4 (2 - q - 2 q^2 - 2 q^3 - q^4 + 2 q^5) u  \nonumber \\ && 
    + 96 m^2 (1 - q - q^2 - q^3 + q^4) u^2 
    +64 (2 - 3 q - 3 q^2 + 2 q^3) u^3 \},
\end{eqnarray}
which is exactly the curve used in \cite{HKK}. The discriminant of the Weierstrass curve $g_2^3-27g_3^2$ is a perfect square. However unlike the theory with fundamental matters, the degenerate roots here in the discriminant do not represent multiple charged massless particles at the discriminant points \cite{HKK}. For later convenience, we will define a new discriminant $\Delta$ without the square and drop some $u$-independent factors 
\begin{eqnarray}
 \Delta= (4u +m^2 ) (4u+ q m^2 q ) (4u + (1+q) m^2 )
 \end{eqnarray}
 
It turns out the normalization of the $\epsilon$ parameter differs by a factor of $2$ from the previous cases. Therefore we propose the following saddle point equation for the $\mathcal{N}=2^*$ theory
\begin{eqnarray} 
q w(x) w(x-\frac{\epsilon}{2}) M(x-\frac{\epsilon}{4}) -w(x)P(x) + 1=0, 
\end{eqnarray}
where $P(x)$ and $M(x)$ are the same as in the quartic curve (\ref{1122.adjointquartic}).

The bare coupling is not renormalized in the massless  $\mathcal{N}=2^*$, or  $\mathcal{N}=4$ theory, but it is renormalized by instanton effects in the $N_f=4$ massless theory. In both theories the bare coupling is renormalized in the massive case.  The elliptic parameter $\tau$ of the Weierstrass curve defined  in (\ref{1111.4.11a}) is the renormalized gauge coupling of the $\mathcal{N}=2^*$ theory. In our convention it has the normalization with respect to the period and prepotential as $\partial_a^2 F^{(0,0)}(a) = -4\pi i \tau$, which differs from the normalization in theories with fundamental matters by a factor of $2$. 

The calculations of the deformed period $\tilde{a}=\sum_{n=0}^{\infty} a_n\epsilon^n$ is similar to the previous cases. We expand $w(x) =\sum_{n=0}^{\infty} w_n(x)\epsilon^n$ and solve $w_n(x)$ recursively by the saddle point equation. We check the leading period $a\equiv a_0 =-\Res_{x=b} x\partial_x\log w_0(x)$  perturbatively  as a series expansion of small $q$ is the same as implied by the relations (\ref{1111.4.11a}, \ref{1111.4.12a}). 

The odd terms in the deformed period vanish. We find the formulae for the non-vanishing even terms $a_2$ and $a_4$. For simplicity we write only $a_2$ formula 
\begin{eqnarray}
a_2 &=& -\frac{1}{96m^2} \{ (2[12u+(1+q)m^2]\partial_u a \nonumber \\ &&
+ [48u^2 +16 (1+q)m^2 u + (1+3q+q^2) m^4]\partial_u^2 a\}
\end{eqnarray}
Similar to the $N_f=4$ theory, we observe that the formula is singular in the massless limit $m\rightarrow 0$, therefore we can not calculate the deformed period directly in the massless limit.

The higher genus formulae are derived in \cite{HKK}, 
\begin{eqnarray}
F^{(1,0)}  &=&  \frac{1}{24}\log(\Delta) - \frac{1}{8} \log(1 - q),  \label{1122.adjF1}\\
F^{(2,0)} &=& \frac{1}{8640\Delta^2 } \{ (37 + 33 q - 39 q^2 - 33 q^3 - 39 q^4 + 33 q^5 + 
          37 q^6) m^{10}  \nonumber \\ && +
       48(12 + 5 q - 12 q^2 - 12 q^3 + 5 q^4 + 12 q^5)  m^8  u  \nonumber \\ &&  
       + 96  (35 - 2 q   - 35 q^2 - 2 q^3 + 35 q^4) m^6 u^2    \nonumber \\ &&   + 
       512 (17 - 9 q - 9 q^2 + 17 q^3)  m^4 u^3
       +  8448 (1 - q + q^2) m^2 u^4   \nonumber \\ &&   
       - 90 [(1 + 3 q + q^2) m^4 + 16  (1 + q) m^2  u + 48 u^2]^2 X\},  \label{1122.adjF2}
\end{eqnarray}
where $X=\frac{E_2(\tau)E_4(\tau)}{E_6(\tau)} \frac{g_3}{g_2}$. Here we add a $\log(1 - q)$ term in the $F^{(1,0)}$ formula besides the discriminant. This term does not affect the equations for deformed dual period, but will be needed for the deformed Matone relation. As similar to the $N_f=4$ theory, this term does not match Nekrasov's instanton partition function, which is 
\begin{eqnarray}
F^{(1,0)}_{Nekrasov}  =  \frac{1}{24}\log(\Delta) - \frac{1}{48} \log\frac{(1 - q)^2q^2}{q_0^2} +\log\theta_3(q_0) 
\end{eqnarray}
We again argue the difference is due to some perturbative contributions. In any case, this subtlety will not appear at higher genus $F^{(n,0)}$ for $n\geq 2$. 

We expand the equation for deformed dual period to higher orders to find differential equations for the higher genus amplitudes 
\begin{eqnarray} 
  \frac{\partial \mathcal{F}(\tilde{a})}{\partial \tilde{a}}  - \tilde{a}_D 
& =&   \frac{\partial F^{(0,0)} (a)}{\partial a } - a_D  
+\epsilon^2 ( \partial_a F^{(1,0)}(a)  -4\pi i \tau a_2  -a_{D2}) \nonumber \\ &&
+\epsilon^4[ \partial_a F^{(2,0)}(a) +a_2  \partial_a^2 F^{(1,0)}(a) -4\pi i \tau a_4 -2 \pi i (\partial_a \tau) (a_2)^2 -a_{D4}] 
 \nonumber \\ &&  +\mathcal{O}(\epsilon^6) 
\end{eqnarray}
The difference with the previous cases (\ref{1111.3.29}, \ref{1118.dual}) is a factor of $2$ in front of $\tau$, due to the different normalization $\partial_a^2 F^{(0,0)}(a) = -4\pi i \tau$ here. We check that our higher genus formulae (\ref{1122.adjF1}, \ref{1122.adjF2}) satisfy the order $\epsilon^2$ and $\epsilon^4$ equations.

We also consider the Matone relation for the $N=2^*$ theory. First after some trials, we write the leading order Matone relation 
\begin{eqnarray} \label{1122.5.11}
q_0\partial_{q_0} F^{(0,0)}_{inst}(a,q_0) -a^2 +\theta_3^4(q_0) u +f_0(q_0) m^2  =0 
\end{eqnarray}
We check this relation perturbatively as series expansion around small $q_0$, using Nekrasov formula for $F^{(0,0)}(a,q_0)$ and the expansion $u(a,q)$ implied by the relations (\ref{1111.4.11a}, \ref{1111.4.12a}). 

The last term in the leading order Matone relation (\ref{1122.5.11}) are independent of period $a$, so they are the integration constant in the formula  $\partial_a^2 F^{(0,0)}(a) = -4\pi i \tau$,  and can not be determined by integrating twice the gauge coupling $\tau$ with respect to $a$. Although this term is not important for higher order equations for the deformed  Matone relation, we can fix it with confidence by computing the Nekrasov partition function to some high instanton numbers as 
\begin{eqnarray}
f_0(q_0) = \frac{4}{3}q_0\partial_{q_0}\log[\frac{\theta_2^2(q_0)\theta_3^2(q_0)}{\theta_4(q_0)}]-\frac{1}{12}
\end{eqnarray}

The classical contribution to prepotential $F_{classical}^{(0,0)}=-\log(q_0) a^2$ absorbs the $-a^2$ term in the Matone relation (\ref{1122.5.11}). We also note that $ \frac{\partial \log(q)} {\partial \log(q_0)} = \frac{\theta_4^4(q_0) }{2}$ according the (\ref{1122.5.2}) and the well known derivative formulae for Jacobi theta functions. So we can write the Matone relation for the total  contribution to prepotential 
\begin{eqnarray} \label{1122.5.13}
q\partial_{q} F^{(0,0)}(a,q)+\frac{2 u}{1-q} +\frac{2f_0(q_0)}{\theta_4^4(q_0)}  m^2  =0 
\end{eqnarray}

We take derivative with respect to $a$ once and twice to find the partial derivatives 
\begin{eqnarray}
\partial_q a_D (a,q) = - \frac{2\partial_a u}{q(1-q)}, ~~~  \partial_q \tau(a,q)  =  \frac{\partial_a^2 u}{2\pi iq(1-q)},
\end{eqnarray} 
They are almost the same as the $N_f=4$ theory except an extra factor of $2$ for $\partial_q a_D$.  Again the partial derivative of $\partial_q u(a,q)$ can be found by taking derivative with respect to $q$ on both sides of (\ref{1111.4.11a}) and use the above formula  for $\partial_q\tau$.

We replace the period and prepotential with the deformed ones in the Matone relation (\ref{1122.5.13}), and expand to higher orders for small $\epsilon$, 
\begin{eqnarray} \label{1122.deformedadj}
 && q\partial_q \mathcal{F}(\tilde{a},q,\epsilon)  +\frac{2u}{1-q}+\frac{2f_0(q_0)}{\theta_4^4(q_0)}  m^2    \\
  && = [q\partial_q F^{(0,0)}(a,q)   +\frac{2u}{1-q}+\frac{2f_0(q_0)}{\theta_4^4(q_0)}  m^2  ] 
    +\epsilon^2 q[ \partial_q F^{(1,0)}(a,q) +a_2 \partial_q a_D] \nonumber \\
&& + \epsilon^4 q [\partial_q F^{(2,0)}(a,q) +a_2 \partial_a\partial_q F^{(1,0)}(a,q) +a_4 \partial_q a_D -  a_2^2 \partial_q (2\pi i \tau) ] +\mathcal{O}(\epsilon^6) \nonumber
\end{eqnarray}
This is almost the same as that of the $N_f=4$ theory (\ref{1119.deformedNf4}) except a factor of $2$ in front of the $\tau$ parameter due to different normalizations. We check our higher genus formulae  (\ref{1122.adjF1}, \ref{1122.adjF2}) satisfy the order $\epsilon^2$ and $\epsilon^4$ equations in the above deformed Matone relation (\ref{1122.deformedadj}).

\section{Derivation of the holomorphic anomaly equation} \label{1214.section6}

In the previous sections, we explicitly check our higher genus formulae satisfy the equations from the saddle point method up to some low genus. It would be nice to directly show the saddle point method is consistent with the holomorphic anomaly equation and gap boundary conditions, and therefore prove the equivalence to all genera. This is considered for the loop equations and topological recursion in matrix models in \cite{EO, EMO}. In this section we will show that under certain simple assumptions,  the holomorphic anomaly equation in the Nekrasov-Shatashvili limit can be derived from the equation $\frac{\partial \mathcal{F}(\tilde{a})}{\partial \tilde{a}} = \tilde{a}_D $  for deformed dual period. 

The generalized holomorphic anomaly equation is proposed in \cite{KW, HK2010, HKK} to solve the higher genus amplitudes of Seiberg-Witten gauge theory in general $\Omega$ background with generic $\epsilon_1, \epsilon_2$ parameters. In the chiral or Nekrasov-Shatashvili limit, the second derivative term in the generalized holomorphic anomaly equation vanishes and the equation is simplified as
\begin{eqnarray} \label{1129.hae}
\partial_{E_2} F^{(n,0)} = \frac{1}{24} \sum_{l=1}^{n-1} \partial_a F^{(l,0)}\partial_a F^{(n-l,0)} 
\end{eqnarray}
Here the amplitude $F^{(n,0)}$ is a polynomial of $X=\frac{E_2(\tau)E_4(\tau)}{E_6(\tau)}\frac{g_3(u)}{g_2(u)}$, and the coefficients of the polynomial are rational functions of $u$. The partial derivative with respect to the second Eisenstein series $E_2$ is  well defined, in the sense by regarding the other components $E_4, E_6$, and $u$ in $F^{(n,0)}$ as constants under the partial derivative. 

The second Eisenstein series $E_2(\tau)$ is holomorphic but not modular under $SL(2,\mathbb{Z})$ transformations. One can instead define a modular covariant but an-holomorphic quantity by a shift $\hat{E}_2(\tau) =E_2(\tau) -\frac{6i}{\pi (\tau-\bar{\tau})}$, which is called an almost holomorphic modular form. The holomorphic limit takes $\bar{\tau}\rightarrow \infty$ and we see $\hat{E}_2(\tau) \rightarrow E_2(\tau)$ in this limit.  It is well known in the theory of modular forms that there is an isomorphism between the almost holomorphic modular forms and the holomorphic limit \cite{Zagier}. 

In the saddle point method we are essentially working in the holomorphic limit where higher genus amplitude is holomorphic but not modular. The holomorphic anomaly appears when we use the isomorphism with almost holomorphic modular forms and replace expression in the holomorphic limit with the almost holomorphic modular counterpart. In our case,  only $E_2$ is not modular covariant and needed to be replaced with the almost holomorphic modular form $\hat{E}_2$. The an-holomorphic derivative can be related to $\partial_{\hat{E}_2}$
\begin{eqnarray} \label{1129.6.2} 
\bar{\partial}_{\bar{\tau}}= (\bar{\partial}_{\bar{\tau}} \hat{E}_2)  \partial_{\hat{E}_2}= \frac{6}{\pi i (\tau-\bar{\tau})^2 } \partial_{\hat{E}_2},
\end{eqnarray} 
which is the origin of the appearance of $\partial_{E_2}$ in holomorphic anomaly equation (\ref{1129.hae}) in the holomorphic limit. 

We will need to derive some formulae involving the operator $\partial_{E_2}$. First we can work in the holomorphic limit for some simple formulae. It is easy to see from the relations (\ref{1111.4.11a}, \ref{1111.4.12a}) that the expressions for the following derivatives have only $E_4, E_6, u$ but no $E_2$, so 
\begin{eqnarray} \label{1129.6.3}
\partial_{E_2}(\partial_a \tau) =0,~~~ \partial_{E_2}(\partial_u a) =0.
\end{eqnarray}
The $E_2$ series starts to appear when we take one more derivative. We can compute 
\begin{eqnarray} \label{1129.6.4}
\partial_{E_2}(\partial^2_u a) =\frac{\pi i}{6} (\partial_u\tau) (\partial_u a)
\end{eqnarray}

We will assume the higher order contributions to the deformed period $\tilde{a}=\sum_{n=0}^{\infty} a_{2n} \epsilon^{2n}$ can be written as a linear combination of $\partial_u a$ and $\partial_u^2 a$. The dual deformed period has the same formula. For $n\geq 1$ we can write 
\begin{eqnarray} \label{1130.6.5}
a_{2n} &=& f_1(u) \partial_u a+f_2(u) \partial_u^2 a,  \nonumber \\
a_{D2n} &=& f_1(u) \partial_u a_D+f_2(u) \partial_u^2 a_D
\end{eqnarray}
where the coefficients $f_1(u)$ and $f_2(u)$ are rational functions of $u$. Using (\ref{1129.6.3}, \ref{1129.6.4}) and $\partial_u a_D =-2\pi i\tau \partial_u a$, we find 
\begin{eqnarray}
\partial_{E_2} (a_{2n}) =\frac{\pi i}{6} (\partial_u\tau) (\partial_u a) f_2(u) = -\frac{1}{12}(2\pi i \tau a_{2n} +a_{D2n}) 
\end{eqnarray}
We obtain formulae for the higher order contributions in the deformed period 
\begin{eqnarray} \label{1129.6.7}
&& \partial_{E_2} (\tilde{a}-a) = -\frac{1}{12} [2\pi i \tau(\tilde{a}-a)+( \tilde{a}_D-a_D)],  \\
&& \partial_{E_2}  [2\pi i \tau(\tilde{a}-a)+( \tilde{a}_D-a_D)] = 0  \label{1129.6.8}
\end{eqnarray}

Our goal is to derive the holomorphic anomaly equation (\ref{1129.hae}) from the equation for the deformed period. We can expand the equation for the deformed period
\begin{eqnarray} \label{1129.6.9}
&& \frac{\partial \mathcal{F}(\tilde{a},\epsilon)}{\partial \tilde{a}} - \tilde{a}_D =0  \nonumber \\
&=& \sum_{n=1}^{\infty} \sum_{k=0}^{\infty} \partial_a^{k+1} F^{(n,0)}(a)  \frac{(\tilde{a}-a)^k}{k!} \epsilon^{2n} 
+ \sum_{k=0}^{\infty} \partial_a^{k+1} (-2\pi i \tau)    \frac{(\tilde{a}-a)^{k+2}}{(k+2)!} \nonumber \\
&& -[2\pi i \tau(\tilde{a}-a)+( \tilde{a}_D-a_D)]  
\end{eqnarray}
where we have separated the prepotential $F^{(0,0)}$ and use the formulae $\partial_a F^{(0,0)}=a_D$ and $\partial_a^2 F^{(0,0)}=-2\pi i \tau$.

The order $\epsilon^2$ and $\epsilon^4$ equations in the above equation (\ref{1129.6.9})  have been written more explicitly before in (\ref{1118.dual}). We can use the equations to compute $\partial_aF^{(n,0)} (a)$ recursively if we have the formulae for $f_1(u)$ and $f_2(u)$ in (\ref{1130.6.5})  for the higher order contributions in the deformed period. Furthermore, by dimensional analysis we know the asymptotic behavior of $F^{(n,0)}\sim a^{2-2n}$ for large $a$. So these equations determine $F^{(1,0)}$ up to a constant and completely fix $F^{(n,0)}$ for $n\geq 2$. 

We would like to derive (\ref{1129.hae}) recursively by induction. Taking the partial derivative $\partial_{E_2}$ on both sides of (\ref{1129.6.9}), we find 
\begin{eqnarray} \label{1129.6.10}
&& \sum_{n=1}^{\infty} \partial_{E_2} \partial_a F^{(n,0)} \epsilon^{2n}  \nonumber \\
&=& -\sum_{n=1}^{\infty} \sum_{k=1}^{\infty} \epsilon^{2n} (\partial_{E_2}\partial_a^{k+1} F^{(n,0)} )\frac{(\tilde{a}-a)^k}{k!} 
-\sum_{n=1}^{\infty} \sum_{k=1}^{\infty} \epsilon^{2n} (\partial_a^{k+1} F^{(n,0)} )\frac{(\tilde{a}-a)^{k-1}}{(k-1)!} \partial_{E_2} (\tilde{a}-a) \nonumber \\ && 
+2\pi i \sum_{k=0}^{\infty} \partial_{E_2}\partial_a^{k+1} \tau  \frac{(\tilde{a}-a)^{k+2}}{(k+2)!} 
+2\pi i \sum_{k=0}^{\infty} \partial_a^{k+1} \tau  \frac{(\tilde{a}-a)^{k+1}}{(k+1)!}  \partial_{E_2}(\tilde{a}-a) ,
\end{eqnarray}
where we have used the equation (\ref{1129.6.8}). At each order $\epsilon^{2n}$, no $F^{(l,0)}$ with $l\geq n$ appears on the right hand side.  So by induction we can use  (\ref{1129.hae}) to compute the right hand side, and we will  complete the induction procedure by showing the left hand side also satisfies the holomorphic anomaly equation  (\ref{1129.hae}). 

It is clear that in order to do the computations, it is crucial to understand how $\partial_{E_2}$ and $\partial_a $ commute with each others. This is mostly conveniently done in the almost holomorphic modular forms, instead of the holomorphic limit.  To preserve the almost holomorphic modular structure, we need to use covariant derivatives with respect to the special Kahler metric of  the moduli space.  There are two contributions to the connection in covariant derivatives, one from the canonical line bundle and one from the Weil-Petersson metric.  In our case, the moduli space  of the Seiberg-Witten theory is similar to that of a one-parameter local Calabi-Yau space, and one can choose a gauge such that the contribution from the canonical line bundle vanishes. So we only need to include the connection from the Weil-Petersson metric. Furthermore, there is a flat coordinate $a$ such that the  connection for the flat coordinate vanishes in the holomorphic limit. The metric and connection in the flat coordinate $a$ in Seiberg-Witten theory are well known, see e.g. \cite{HK2009}, 
\begin{eqnarray}
G_{a\bar{a}} \sim (\tau-\bar{\tau}),~~~ \Gamma_{aa}^a=(G_{a\bar{a}})^{-1}\partial_a G_{a\bar{a}} =\frac{\partial_a \tau }{\tau-\bar{\tau} } 
\end{eqnarray}
where we see the Christoffel connection indeed vanishes in the holomorphic limit $\bar{\tau}\rightarrow \infty$. 

Suppose $F_k$ is a tensor with $k$ lower indices regarding to the metric of the moduli space in flat coordinate $a$, and it may has an-holomorphic dependence in terms of $\hat{E}_2$. The covariant derivative is then $D_a F_k =(\partial_a -k \Gamma_{aa}^a) F_k$.  We can compute the an-holomorphic derivative 
\begin{eqnarray}
\bar{\partial}_{\bar{\tau}} D_a  F_k = (\partial_a -k \Gamma_{aa}^a) \bar{\partial}_{\bar{\tau}} F_k  -k ( \bar{\partial}_{\bar{\tau}} \Gamma_{aa}^a ) F_k
\end{eqnarray} 
We use (\ref{1129.6.2}) and then take the holomorphic limit to find the commutation relation 
\begin{eqnarray}
\partial_{E_2}\partial_a F_k =\partial_a \partial_{E_2}  F_k -\frac{k\pi i}{6} (\partial_a \tau) F_k
\end{eqnarray}

The amplitude $F^{(n,0)} $ is a scalar in moduli space, and its derivative with $\partial_a$ is a tensor with lower indices. We can compute the derivatives 
\begin{eqnarray} \label{1129.6.14}
\partial_{E_2}\partial_a^{k+1} F^{(n,0)} &=& \partial_a^{k+1} \partial_{E_2} F^{(n,0)} -\frac{\pi i}{6} \sum_{l=1}^k l\partial_a^{k-l} [\partial_a \tau \partial_a^l F^{(n,0)} ] \nonumber \\
&=&  \partial_a^{k+1} \partial_{E_2} F^{(n,0)} -\frac{\pi i }{6} \sum_{p=0}^{k-1} \binom{k+1}{p+2} (\partial_a^{p+1} \tau) (\partial_a^{k-p}F^{(n,0)} ), 
\end{eqnarray} 
where we have used the binomial identity $\sum_{l=1}^{k-p} \binom{k-l}{p} l= \binom{k+1}{p+2}$. In particular, we note that in the case of $k=0$, the operators $\partial_{E_2}$ and $\partial_a$ commute when acting on  $F^{(n,0)} $. 

Similarly we derive the  formula for $\tau=-\frac{1}{2\pi i} \partial_a^2 F^{(0,0)}$, using the first formula in (\ref{1129.6.3}) 
\begin{eqnarray} \label{1129.6.15}
\partial_{E_2}\partial_a^{k+2}\tau = -\frac{\pi i }{12} \sum_{p=0}^{k}\binom{k+4}{p+2}  (\partial_a^{p+1}\tau)( \partial_a^{k+1-p}\tau )
\end{eqnarray}

Further using the equation for deformed dual period (\ref{1129.6.9}), the formula (\ref{1129.6.7}) can be written without the dual period as 
\begin{eqnarray} \label{1129.6.16}
\partial_{E_2} (\tilde{a}-a) &=&  -\frac{1}{12}  \sum_{n=1}^{\infty} \sum_{k=0}^{\infty} \partial_a^{k+1} F^{(n,0)}(a)  \frac{(\tilde{a}-a)^k}{k!} \epsilon^{2n} 
\nonumber \\ && +\frac{\pi i}{6}  \sum_{k=0}^{\infty} (\partial_a^{k+1}  \tau)    \frac{(\tilde{a}-a)^{k+2}}{(k+2)!}
\end{eqnarray}

We can now compute the right hand side of (\ref{1129.6.10}), by plugging  the formulae (\ref{1129.6.14}, \ref{1129.6.15}, \ref{1129.6.16}) and then use (\ref{1129.hae}) by induction. The calculation is quite lengthy, but surprisingly we encounter a lot of cancellations which drastically simplify the expression. In particular, the dependence on $(\tilde{a}-a)$ cancels out, so we don't need the specific formulae for $f_1(u)$ and $f_2(u)$ in (\ref{1130.6.5}). We keep the left hand side of (\ref{1129.6.10}) and write the final result of the calculations for the right hand side
\begin{eqnarray} 
\sum_{n=1}^{\infty} \partial_a  \partial_{E_2} F^{(n,0)} \epsilon^{2n} = \frac{1}{12} \sum_{n=1}^{\infty} \epsilon^{2n} \sum_{l=1}^{n-1} \partial_aF^{(l,0)} \partial_a^2 F^{(n-l,0)} 
\end{eqnarray}

It is easy to check $\partial_{E_2} F^{(1,0)} =0$, thus the above result proves the holomorphic anomaly equation (\ref{1129.hae}) for $F^{(n,0)}$ with $n\geq 2$ up to an integration constant of $a$. From the asymptotic behavior $F^{(n,0)}\sim a^{2-2n}$ for large $a$, the constant must be zero, so we have proven (\ref{1129.hae}) exactly by induction.

After the successful derivation of the holomorphic anomaly equation from the equation for the deformed period, one may wonder whether it can be also derived from the deformed Matone relation. However, there is one important difference between these two equations. We have noted that the  equation for the deformed period determines $\partial_a F^{(n,0)}$ recursively  and the asymptotic behavior $F^{(n,0)}\sim a^{2-2n}$ further fix the integration constant at $a\sim \infty$ to be zero for $n\geq 2$. On the other hand, the deformed Matone relation  determines $\partial_q F^{(n,0)}$ recursively, and the integration constant here is the perturbative contribution to $F^{(n,0)}$ at $q=0$, which is independent of the instanton counting parameter $q$ for $n\geq 2$. The perturbative contribution is crucial in summing together with the instanton contributions into our higher genus formulae. Without this piece of information, we expect it is difficult to derive the holomorphic anomaly equation (\ref{1129.hae}) or the gap boundary condition from the deformed Matone relation.

\section{Topological string theory on local Calabi-Yau manifolds}

The refined topological string invariants with two expansion parameters $\epsilon_1$ and $\epsilon_2$ can jump in the complex structure moduli space, and are in general difficult to study. However on certain local toric Calabi-Yau manifolds where there is no complex structure deformation, the refined topological string amplitudes can be computed by A-model method of the refined topological vertex \cite{IKV}, or by the mirror B-model method with a generalized holomorphic anomaly equation and the gap boundary conditions \cite{HK2010}.  In this section we consider applying the techniques developed in earlier sections for $SU(2)$ Seiberg-Witten theory  to topological string theory on some local Calabi-Yau manifolds, in the chiral or Nekrasov-Shatashvili limit where one of the $\epsilon$ parameters vanishes.  Here there will be no analog of deformed Matone relation as in the Seiberg-Witten theory, and we will only consider the equation for deformed dual period 
\begin{eqnarray}  \label{1214.7.1}
\partial_{\tilde{t}} \mathcal{F}(\tilde{t},\epsilon) =\tilde{t}_D
\end{eqnarray} 
where $\mathcal{F}(\tilde{t},\epsilon)=\sum_{n=0}^{\infty} F^{(n,0)}\epsilon^{2n}$ is the higher genus refined amplitudes in the Nekrasov-Shatashvili limit, and $\tilde{t}$ and $\tilde{t}_D$ are the deformed period and dual period whose leading order contributions in small $\epsilon$ are the usual period and dual period. 

The Nekrasov-Shatashvili limit of the refined topological string theory has been considered in \cite{ACDKV}. The novelties here are the following points. We will derive exact formulae for the higher order contributions to the deformed period and dual period. Together with the equation for the deformed period (\ref{1214.7.1}), these formulae enable us to write differential equations for the higher genus amplitudes $F^{(n,0)}(t)$. The differential equations compute $\partial_t F^{(n,0)}$ recursively and determine $F^{(n,0)}$ up to a constant. We can then check the higher genus formulae in \cite{HK2010} satisfy these differential equations exactly to all degrees of world sheet instanton. Furthermore, similar to the gauge theory case, we can show that these differential equations imply the generalized holomorphic anomaly equation in Nekrasov-Shatashvili limit, thus taking another step toward elucidating the mirror symmetry between the A-model and B-model.  

The topological string amplitudes on Calabi-Yau manifolds have two contributions when we expand around the large volume point in the moduli space, the constant map contributions and the world sheet instanton contributions.  The constant map contribution in conventional topological string theory at a given genus is a constant related to the Bernoulli numbers, and has been computed in \cite{FP, GV, Marino}. The refined version of the constant map is not quite clear here, and it is not determined in the A-model and B-model methods in \cite{IKV, HK2010} either. We will be agnostic about constant map contribution here as well and our equations only fix the world sheet instanton contributions which vanishes in the large volume limit $t\sim \infty$.

The study for the topological string case is similar to that of the gauge theory case, with the exception of two technical points. Firstly, the modular group generated by the monodromy around the special points in the moduli space is in general not a subgroup of $SL(2,\mathbb{Z})$, so our formalism for $SU(2)$ Seiberg-Witten theory in terms of Eisenstein series would not be available. We will define certain (almost) modular generators, and the higher genus topological string amplitudes and their derivatives can be written as rational functions of these generators. We will need to find the derivative rules for these generators, in place of the well known Ramanujan derivative identities for the Eisenstein series. Although the formalism in terms of Eisenstein series and Jacobi theta functions is still available for certain special Calabi-Yau models, such as the local $\mathbb{P}^2$ model, discussed in \cite{ABK}, we will not resort to the formalism for the sake of generality. 

Secondly, the period and dual period are a power series and a log series in the case of $SU(2)$ Seiberg-Witten theory, but they are a log series and a double log series in the case of topological string on local Calabi-Yau manifolds. This can be easily seen in their respective Picard-Fuchs differential equations satisfied by the period and dual period. We can compute the period perturbatively by a contour integral or residue calculations. It turns out the leading logarithmic term does not appear in the residue calculations of leading order period, and needed to be added manually. For the higher order contributions to the exact deformed period, the logarithmic term will resurface in the residue  calculations.

\subsection{Local $\mathbb{P}^2$ model}

The local Calabi-Yau 3-fold is a complex line bundle over $\mathbb{CP}^2$, and is one of most well studied models, see e.g. \cite{HKR}. The local mirror geometry has one complex structure modulus parameter $z$ and can be reduced to the following curve
\begin{eqnarray} \label{1214.7.2}
H(x,p)=-1+e^x +e^p -z e^{\epsilon /2}e^{-x}e^{-p} =0,
\end{eqnarray}
which is also used in \cite{ACDKV} with a minus sign conventional difference for the parameter $z$.   The mirror curve can be treated as a quantum mechanical Hamiltonian and the coordinates $x$ and $p$ are the conjugate parameters for position and momentum. We quantize the curve by imposing noncommutativity relation $[x,p]= i\hbar$ and use the notation $\epsilon\equiv -i\hbar$. We should note that the factor $e^{\epsilon /2}$ in the last term in the curve (\ref{1214.7.2}) does not appear at the classical level, but emerges at the quantum level to compensate for the noncommutativity of $e^{-x}$ and $e^{-p}$ in the last term. 

The wave function of the quantum mechanical Hamiltonian can be written as 
\begin{eqnarray} \label{1214.7.3}
\psi(x)= \exp (\frac{1}{\epsilon} \int^{x} w(x) dx )
\end{eqnarray}
We are interested in the eigenstate with zero energy $H\psi(x) =0$, and we will solve for $w(x)$ perturbatively around small $\epsilon$ parameter in the WKB approximation. To do this we should first understand how $H$ and in particular $e^{p}$ act on the wave function. The canonical representation of the momentum operator in quantum mechanics which fulfills the noncommutativity relation is $p=-i\hbar \partial_x = \epsilon \partial_x$. It is easy to see 
\begin{eqnarray}
e^{p} \psi(x) = \psi(x+\epsilon), ~~~ e^{-p} \psi(x) = \psi(x-\epsilon)
\end{eqnarray}

The function $w(x)$  appearing in the exponent in the wave function(\ref{1214.7.3}) has implicit dependence on $\epsilon$ parameter.   We expand the $w(x)$ function as 
\begin{eqnarray}
w(x) = \sum_{n=0}^{\infty} w_0(x) \epsilon^n
\end{eqnarray}
where $w_n(x)$ is independent of $\epsilon$ parameter. The conventional period $t$ and dual period $t_D$ of the mirror geometry are computed by the contour integral of the leading term as $\oint w_0(x)dx$. The period $t$ is the flat coordinate whose connection vanishes in the holomorphic limit and its exponential $Q=e^t\sim z$ is the A-model expansion parameter in the large volume point $z\sim 0$ in our parametrization. We refer to the contour integral of $w(x)$ including the higher order contributions as the deformed or the quantized (dual) period, denoted as by tilde symbol as $\tilde{t}$ and $\tilde{t}_D$.   

We can then expand the Schrodinger equation $H\psi(x)=0$ for small $\epsilon$ and at each order $\epsilon^n$ we find equation for $w_n(x)$ in terms of lower order terms.  At the leading order the $w_0(x)$ can be obtained by solving for $p$ in terms of the $x$ in the curve (\ref{1214.7.2}) at the classical level. There are two solutions to the quadratic equation and their contour integrals have opposite signs. We can choose the one with the right convention  
\begin{eqnarray} 
w_0(x) =\log [- \frac{1}{2} (e^x-1 + \sqrt{ (e^x-1)^2 + 4 ze^{-x} })~] 
\end{eqnarray}
The higher order functions $w_n(x)$ can be solved recursively, and we list them up to a few orders 
\begin{eqnarray}
w_1(x) &=& \frac{ - e^{3 x} +e^{2 x} + 2 z}{2 e^x(e^x-1)^2  + 8 z}  \nonumber \\
w_2(x) &=& -e^{\frac{x}{2}}  [2 e^{6 x} -6 e^{5 x} + e^{4 x} (6 - 99 z)  +e^{3 x} (-2 + 157 z)  - 69 e^{2 x} z   \nonumber \\ && + e^x z (11 + 144 z)  - 16 z^2] /[ 24 (e^x(e^x -1)^2 + 4 z)^{5/2}]
 \\ \nonumber &&  \cdots 
\end{eqnarray}

We can compute the contour integrals perturbatively around the large volume point $z\sim 0$. The expansion of $w_n(x)$ around small $z$ gives rise to a series whose coefficients are rational function of $e^x$. The rational functions have a pole at $e^x-1$ or $x=0$. It is straightforward to compute the residue around $x=0$, and it turns out that the period corresponds to the residue at $x=0$. On the other hand, the dual period  corresponds to the integral of a more complicated contour, but since many equations for the period are also valid for the dual period, we will directly use these same equations and will not need to do the more complicated contour integral here.  

We denote the quantized (dual) period in terms of the expansion 
\begin{eqnarray}
\tilde{t} =\sum _ {n=0}^{\infty} t_{n}\epsilon^n ,   ~~~~ \tilde{t}_D =\sum _ {n=0}^{\infty} t_{Dn} \epsilon^n,
\end{eqnarray}
where the leading order terms are also denoted as $t\equiv t_0$ and $t_D\equiv t_{D0}$. The residue of $w_0(x)$ around $x=0$ provides the power series in the leading period $t$, and after including the correct factor of $3$ and add the leading logarithmic term, we can write the period as 
\begin{eqnarray} \label{1214.7.8}
t &=&  \log(z) +\frac{3}{2\pi i}\oint_{x=0}  w_0(x) dx \nonumber \\
&=& \log(z) -6 z + 45 z^2 - 560 z^3 + \frac{17325 z^4}{2} - \frac{756756 z^5}{5}+\mathcal{O}(z^6) 
\end{eqnarray} 

The exact series for leading period $t$  are characterized by the Picard-Fuchs equation $\mathcal{D}t=0$, where the operator is
\begin{eqnarray}
\mathcal{D} = \Theta_z ^3 + 3z (3\Theta_z+2)(3\Theta_z+1)\Theta_z,
\end{eqnarray}
with the notation  $\Theta_z= z\partial_z$.  The leading dual period $t_D$ has a double logarithmic leading term $(\log z)^2$ and also satisfies the same Picard-Fuchs equation $\mathcal{D}t_D=0$. 

We can compute the higher order contributions to the quantized (dual) period in (\ref{1214.7.8}) by the residue 
\begin{eqnarray}
t_n =\frac{3}{2\pi i} \oint_{x=0} w_n(x) dx,  ~~~ n\geq 1
\end{eqnarray} 
Here comparing with the case of $n=0$ in (\ref{1214.7.8}),  we do not need to manually include an extra leading term besides the contour integral. For an odd integer $n$, the integrand $w_n(x)$ can be written as a total derivative of simple functions. For the case of $n=1$, the total derivative is a logarithmic function, and there is a rather trivial contribution of a constant to the residue $T_1=-\frac{3}{2}$. At higher orders with odd $n>1$, the total derivative is a rational function of $e^x$, so there is no branch cut and the residue vanishes $t_n=0$. Since the dual period has the same integrand, albeit a more complicated contour, the odd terms in the quantized dual period also vanish $t_{Dn}=0$, with possibly the exception of a trivial contribution at the first order $n=1$. The rather trivial contributions at $n=1$ will not affect our formalism. In the followings we will only need to consider the even power terms.   

Similar to Seiberg-Witten theory, we find the non-vanishing even higher order contributions to the quantized period $t_{2n}$ can always be written as a linear combination of the first and second derivatives of the leading period $t$. Specifically, we find the exact formulae 
\begin{eqnarray} \label{1214.formulae}
t_2 &=& \frac{\Theta_z^2 t }{8} ,   \nonumber  \\ 
t_4 &=& \frac{2z (999 z -5) \Theta_z t + 3 z (2619 z -29) \Theta_z^2 t}{640 \Delta^2},  \\
t_6 &=& \frac{z}{107520  \Delta ^4} [2 (25690689 z^3 -3140937 z^2 +29031 z -7 ) \Theta_z t  \nonumber \\  && 
+  (176694291 z^3 -27479655 z^2 +363285 z -137   ) \Theta_z^2 t] , \nonumber 
\end{eqnarray}
where $\Theta_z=z\partial_z $ and $\Delta =1+27 z$ is the discriminant of the mirror geometry. 

These exact formulae (\ref{1214.formulae}) are derived by showing the integrands in the relevant contour integrals can be written as  total derivatives  with respect to $x$. It turns out that there is a logarithmic piece in the total derivative, and the residue around $x=0$ does not completely vanish. Detailed calculations show that the contributions from this logarithmic branch cut exactly accounts for the leading $\log(z)$ term in formula (\ref{1214.7.8}) for the leading period $t$. We also check these formulae  (\ref{1214.formulae}) perturbatively by computing the relevant residues around $x=0$ as series expansions around small $z$.   

The dual period is defined by the same integrand as the period, albeit a more complicated contour. The above arguments also works for the dual period, so the exact formulae (\ref{1214.formulae}) are also valid for the dual period by simply replacing $t$ with $t_D$ in the formulae. 

It is well known that the prepotential is determined by the equation $\partial_t F^{(0,0)}(t) = t_D$. Generalizing the equation to the quantum version, we can derive differential equations for the higher genus amplitudes 
\begin{eqnarray} \label{1214.7.13}
&&\partial_{\tilde{t}} \mathcal{F}^(\tilde{t},\epsilon) - \tilde{t}_D =0 \nonumber \\
&=& \partial_t F^{(0,0)}(t)  -  t_D
+\epsilon^2 [ \partial_t F^{(1,0)}(t)  + t_2 \partial_t^2 F^{(0,0)}(t)   - t_{D2}]
+\epsilon^4~ [\partial_t F^{(2,0)}(t) \nonumber \\ &&  +t_2  \partial_t^2 F^{(1,0)}(t)  +  \frac{(t_2)^2}{2}  \partial_t^3 F^{(0,0)}(t) + t_4  \partial_t^2 F^{(0,0)}(t)  -t_{D4}] 
 +\mathcal{O}(\epsilon^6) 
\end{eqnarray}
At each order $\epsilon^{2n}$, we find an equation for the $\partial_tF^{(n,0)}(t)$ in terms of lower genus amplitudes and the higher order contributions to the quantized (dual) period.

The equations can be simplified a little more by eliminating the dual period.  We act  the operator $\Theta_z$ once and twice on the both sides of the leading order equation $ t_D =\partial_t F^{(0,0)}(t) $,  and find 
\begin{eqnarray}
\Theta_z t_D &=& \partial_t^2 F^{(0,0)}(t)   \Theta_z t, \nonumber \\
\Theta_z^2 t_D &=& \partial_t^2 F^{(0,0)}(t)   \Theta_z^2 t +  \partial_t^3 F^{(0,0)}(t)   (\Theta_z t)^2 
\end{eqnarray}
We notice the dual period only appears at the order $\epsilon^{2n}$ equation as in the combination $t_{2n}  \partial_t^2 F^{(0,0)}(t)  -t_{D2n} $. Suppose at order $\epsilon^{2n}$ we have the exact formula for the period and dual period 
\begin{eqnarray}
t_{2n} &=& x_1  \Theta_z t + x_2  \Theta_z^2 t,   \nonumber \\
 t_{D2n} &=& x_1  \Theta_z t_D + x_2  \Theta_z^2 t_D,
\end{eqnarray}
where $x_1$ and $x_2$ are some rational functions of $z$. We can compute 
\begin{eqnarray} \label{1215.7.16}
t_{2n}  \partial_t^2 F^{(0,0)}(t)  -t_{D2n} = -x_2   \partial_t^3 F^{(0,0)}(t)   (\Theta_z t)^2 
\end{eqnarray}
So we can eliminate the dual period at each higher order in the equation (\ref{1214.7.13}), and express $\partial_tF^{(n,0)}(t)$ in terms of lower genus amplitudes, the derivatives $\Theta_z t$ and $\Theta_z^2 t$  and some rational functions of $z$ which appear in the exact formulae (\ref{1214.formulae}) for the higher order contributions to the quantized period. 

One can already use these equations (\ref{1214.7.13}) to compute $F^{(n,0)}$ recursively as perturbative series around the large volume point $z\sim 0$, utilizing the asymptotic expansion of period $t$ around this point  (\ref{1214.7.8}) and also the formulae (\ref{1214.formulae}) for higher order contributions. The results can be compared with the A-model calculations by the refined topological vertex, as considered in \cite{ACDKV}. We would like to go a step further and check the higher genus formulae in \cite{HK2010} exactly in all orders of the small $z$ expansion. 

The higher genus formulae in \cite{HK2010} are derived by holomorphic anomaly equation and the gap boundary conditions near the conifold point. We quote the formulae in Nekrasov-Shatashvili limit up to genus 3 
\begin{eqnarray} \label{1214.higher}
 F^{(1,0)} &=&  \frac{1}{24}\log(\frac{\Delta}{z}),  \\
 F^{(2,0)}&=& \frac{10 S+(1296 z+11) z^2}{11520 z^2 \Delta^2},  \nonumber \\
 F^{(3,0)}&=& \frac{1}{69672960 z^6 \Delta^4} [280 S^3+420 S^2 (108 z-1) z^2+42 S \left(209952 z^2-4212 z+5\right) z^4
\nonumber \\ && +\left(1167753024 z^3-29387448 z^2+355536
   z+2269\right) z^6 ] \nonumber 
\end{eqnarray}
where $\Delta=1+27z$ is the discriminant and $S$ is an almost holomorphic generator similar to the shifted second Eisenstein series in Seiberg-Witten theory, and its holomorphic limit is an almost modular form. 

We review some formulae for the generator $S$ and special geometry for the local $\mathbb{P}^2$ model, which can be found in \cite{HK2010} and derived in details in \cite{HKR}.  In the holomorphic limit,  the metric in flat coordinate $t$ is a constant up to an anti-holomorphic factor. So the metric and Christoffel connection in the moduli space in the $z$ coordinate are 
\begin{eqnarray} \label{1215.7.18}
G_{t\bar{t}}\sim 1, ~~~ G_{z\bar{z}} =|\partial_z t|^2 G_{t\bar{t}} \sim \partial_z t,~~~ \Gamma_{zz}^z = G^{z\bar{z}}\partial_z G_{z\bar{z}} = \partial_t z\partial_z^2 t
\end{eqnarray}
The three point Yukawa coupling is 
\begin{eqnarray}
C_{zzz} =D_zD_z D_z F^{(0,0)} =   -\frac{1}{3z^3(1+27z)}
\end{eqnarray}
The generator $S\equiv S^{zz}$ is a tensor, also known as the propagator, which satisfies the following relations  
 \begin{eqnarray}
&& \Gamma^z_{zz} = -C_{zzz}S-\frac{7+216z}{6z\Delta}, \nonumber \\
&& D_z S =\partial_zS +2\Gamma_{zz}^z S = -C_{zzz}(S)^2-\frac{z}{12\Delta} 
\end{eqnarray}
So the second derivative of flat coordinate $t$ can be expressed in terms of first derive and the propagator $S$, 
\begin{eqnarray}
 \partial_z^2 t =-(C_{zzz}S+\frac{7+216z}{6z\Delta})   (\partial_z t)^2
 \end{eqnarray}
 
We can express everything as rational functions of three independent generators, for example we can choose $z$, $\partial_tz$ and $S$ as independent generators. The derivatives of the three generators can be again expressed as rational functions of themselves, similarly as in the Ramanujan derivative identities for the Eisenstein series.  So we can compute the higher derivatives of higher genus formulae in terms of the  three independent generators. As for the genus zero case, we note that after we eliminate the dual period with (\ref{1215.7.16}), only the derivatives $\partial_t^k F^{(0,0)}$ with $k\geq 3$ appear in the equations in (\ref{1214.7.13}), so we can start with the three point Yukawa coupling $\partial_t^3 F^{(0,0)} =(\partial_t z)^3 C_{zzz}$ and compute higher derivatives recursively. 
 
Utilizing these derivative relations, we check our higher genus formulae (\ref{1214.higher}) satisfy the differential equations exactly to all orders in $z$ parameter, and up to order $\epsilon^6$ in (\ref{1214.7.13}). 

We expect the derivation of the holomorphic anomaly from the equations in (\ref{1214.7.13}) works similarly as in the Seiberg-Witten theory in Section.\ref{1214.section6}, with the generator $S$ plays the role of $E_2$ there. Since $S$ is the only an-holomorphic generator, the simplified holomorphic anomaly equation in the Nekrasov-Shatashvili limit is 
\begin{eqnarray} \label{1215.hae}
(\partial_t z)^2 \frac{\partial F^{(n,0)}}{\partial S} = \frac{1}{2} \sum_{l=1}^{n-1} \partial_tF^{(l,0)}\partial_tF^{(n-l, 0)}
\end{eqnarray}
Since everything can be written as rational functions of three generators $z, \partial_t z$ and $S$, the partial derivative with respect to $S$ is well defined by treating the other two holomorphic generators $z$ and $\partial_t z$ as constants under the partial derivative. 

We will not go into further details of the lengthy calculations other than working out the commutation relation of $\partial_S$ and $\partial_t$ acting on a tensor. This  commutation relation is crucial for the proof of holomorphic anomaly equation in Section.\ref{1214.section6} from the equation for deformed period. Again to derive the relation we should work with the almost holomorphic modular structure. First we note that the propagator $S$ is  defined by its anti-holomorphic derivative  $\bar{\partial}_{\bar{z}} S =  \bar{C}_{\bar{z}}^{zz} $, 
where $   \bar{C}_{\bar{z}}^{zz}  = \bar{C}_{\bar{z}\bar{z}\bar{z}} (G_{z\bar{z}})^{-2}e^{2K} $ is related to the complex conjugate of the three point Yukawa coupling. So the anti-derivative is related to the partial derivative $\partial_S$ by
\begin{eqnarray} \label{1215.7.23}
\bar{\partial}_{\bar{t}} =\bar{\partial}_{\bar{t}}S  \partial_S = (\bar{\partial}_{\bar{t}} \bar{z} ) \bar{C}_{\bar{z}}^{zz}    \partial_S = (\partial_t z)^2 \bar{C}_{\bar{t}}^{tt}   \partial_S
\end{eqnarray}

In local Calabi-Yau geometry we can choose a gauge such that the holomorphic derivative of the Kahler potential $K$ is trivial. The well known special geometry relation for the moduli space in flat coordinate is simplified for the local case in the holomorphic limit 
\begin{eqnarray}
\bar{\partial}_{\bar{t}}(\Gamma_{tt}^t) = -   C_{ttt}  \bar{C}_{\bar{t}}^{tt}  
\end{eqnarray}

Suppose $F_k$ is a tensor with $k$ lower indices in the flat coordinate $t$. We act the anti-holomorphic derivative 
$\bar{\partial}_{\bar{t}}$ on the covariant derivative $D_t F_k =\partial_t F_k -k\Gamma_{tt}^t F_k$ and find 
\begin{eqnarray} \label{1217.7.26}
 \bar{\partial}_{\bar{t}}D_t F_k  = D_t \bar{\partial}_{\bar{t}}F_k  -  k(\bar{\partial}_{\bar{t}}\Gamma_{tt}^t )F_k =D_t \bar{\partial}_{\bar{t}}F_k  +  k C_{ttt}  \bar{C}_{\bar{t}}^{tt}      F_k 
 \end{eqnarray}
We plug the equation (\ref{1215.7.23}) into the above equation (\ref{1217.7.26}) and then take the holomorphic limit to cancel out the anti-holomorphic factor $\bar{C}_{\bar{t}}^{tt} $. We find the commutation relation 
\begin{eqnarray} \label{1215.commutation}
(\partial_t z)^2 \partial_S\partial_t F_k =\partial_t[(\partial_t z)^2 \partial_SF_k ] + (kC_{ttt}) F_k 
\end{eqnarray}

We apply the commutation relation to $\partial^k_t F^{(n,0)}$ for higher genus amplitudes with $n\geq 1$, and $\partial^k_t C_{ttt} =\partial_t^{k+3} F^{(0,0)}$ for the genus zero amplitude. We obtain the analog of formulae (\ref{1129.6.14}, \ref{1129.6.15}) in Seiberg-Witten theory 
\begin{eqnarray} \label{1215.7.29} 
(\partial_t z)^2 \partial_{S}\partial_t^{k+1} F^{(n,0)} 
&=&  \partial_t^{k+1} [ (\partial_t z)^2 \partial_{S} F^{(n,0)}] + \sum_{p=0}^{k-1} \binom{k+1}{p+2} (\partial_t^p C_{ttt}) (\partial_t^{k-p}F^{(n,0)} ), \nonumber \\
 (\partial_t z)^2 \partial_{S}\partial_t^{k+1}C_{ttt} &=&  \frac{1 }{2} \sum_{p=0}^{k}\binom{k+4}{p+2}  (\partial_t^{p}C_{ttt})( \partial_t^{k-p}C_{ttt}  )
\end{eqnarray}
where $n\geq 1$ in the first formula, and we have used the holomorphicity of the three point coupling $\partial_S C_{ttt}=0$ in the second formula. We check the formulae (\ref{1215.7.29}) explicitly up to some finite integer $k$, using the higher genus formulae (\ref{1214.higher}) and the derivative relations between the three generators. Utilizing these formulae (\ref{1215.7.29}), it is then straightforward to derive the simplified holomorphic anomaly equation (\ref{1215.hae}) by induction from the equation for quantized dual period (\ref{1214.7.13}).  
 
 We should mention that the first formula in (\ref{1215.7.29}) is valid even if we replace the higher genus amplitude $F^{(n,0)}$ with any rational function of the two generators $S$ and $z$, but without $\partial_t z$. This is because  any rational function of the two generators $S$ and $z$ is a scalar and modular invariant, so our arguments still apply.

\subsection{Local $\mathbb{P}^1\times \mathbb{P}^1$ model}
The topological string amplitudes on local $\mathbb{P}^1\times \mathbb{P}^1$ Calabi-Yau is equivalent to the Nekrasov function for the supersymmetric 5-dimensional $SU(2)$ Yang-Mills theory compactified on a circle. It turns out that although this model has two Kahler parameters, it is still somewhat similar to a one-parameter model due to the symmetry between the two parameters. The study is therefore similar to the local $\mathbb{P}^2$ model in the previous subsection, and we would not need to write too much details which have been described before. 

The mirror curve for the model is 
\begin{eqnarray}
H(x,p) = -1 +e^x+e^p +z_1 e^{-x} +z_2e^{-p} =0 
\end{eqnarray}  
where $z_1$ and $z_2$ are the two complex structure parameters. The mirror map to the Kahler parameters $T_1\sim \log(z_1)$ and $T_2\sim \log(z_2)$ near the large volume point $z_1=z_2=0$ are given by the solutions to the Picard-Fuchs equations $\mathcal{L}_1f= \mathcal{L}_2f=0$ with the following  operators
\begin{eqnarray} \label{1222.PF}
\mathcal{L}_1&=& \Theta_1^2 -2z_1(\Theta_1+\Theta_2)(1+2\Theta_1+\Theta_2),  \nonumber \\
\mathcal{L}_2 &=& \Theta_2^2 -2z_2(\Theta_1+\Theta_2)(1+2\Theta_1+\Theta_2), 
\end{eqnarray}
where $\Theta_i=z_i\frac{\partial}{\partial z_i}$,  $i=1,2$.  The discriminant is $z_1z_2\Delta=0$ where
\begin{eqnarray}
\Delta=1-8(z_1+z_2)+16(z_1-z_2)^2
\end{eqnarray}

The first few orders expansion for $T_1$ and $T_2$ are 
\begin{eqnarray}
T_1 &=& \log(z_1) + 2(z_1+z_2) +3(z_1+4z_1z_2+z_2^2)+ \mathcal{O}(z^3), \nonumber \\
T_2 &=& \log(z_2) + 2(z_1+z_2) +3(z_1+4z_1z_2+z_2^2)+ \mathcal{O}(z^3)  
\end{eqnarray} 
We see the power series in $T_1$ and $T_2$ are the same, which can be easily confirmed by checking $T_1-T_2=\log(z_1)-\log(z_2)$ is also a solution to the Picard-Fuchs equations with the operators (\ref{1222.PF}). For convenience, we will use the variables instead
\begin{eqnarray}
t\equiv t_{+}= \frac{T_1+T_2}{2},~~~~ t_{-}= \frac{T_1- T_2}{2} = \frac{\log(z_1)-\log(z_2)}{2} 
\end{eqnarray}

There are four linearly independent solutions to the Picard-Fuchs equations. Besides the constant solution, $T_1$ and $T_2$, we denote the fourth solution as $t_D$ and it has the double logarithmic asymptotic behavior $t_D\sim \log(z_1)\log(z_2)$. It is related to the prepotential by the differential equation 
\begin{eqnarray} \label{1222.df}
t_D= \frac{\partial F^{(0,0)}(T_1,T_2)}{\partial T_1}  +\frac{\partial F^{(0,0)}(T_1,T_2)}{\partial T_2}
= \frac{\partial F^{(0,0)}(t, t_{-})}{\partial t} ,       
\end{eqnarray}
which can be checked by A-model calculations by topological vertex \cite{AKMV}. The prepotential $F^{(0,0)}$ consists of the world-sheet instanton contributions and perturbative contribution. The world-sheet instanton contributions are positive powers of $Q_1=e^{T_1}$ and $Q_2=e^{T_2}$, while the perturbative contribution is the cubic polynomial
\begin{eqnarray}
F^{(0,0)}_{pert} =\frac{1}{24} (T_1^3-3T_1^2T_2-3T_1T_2^2 +T_2^3) 
\end{eqnarray} 

As usual we generalize the differential equation (\ref{1222.df}) to quantum version 
\begin{eqnarray} \label{1222.qmdf}
\tilde{t}_D = \frac{\partial \mathcal{ F}(\tilde{t}, t_{-},\epsilon)}{\partial \tilde{t}}
\end{eqnarray}
where the deformed prepotential is related to the higher genus amplitudes, and we have also  replaced the periods by their quantum deformations 
\begin{eqnarray}
\mathcal{ F}(t, t_{-},\epsilon)= \sum_{n=0}^{\infty} F^{(n,0)}(t,t_{-}) \epsilon^{2n}, ~~~~ \tilde{t} =\sum_{n=0}^{\infty} t_{n}  \epsilon^{n} , ~~~~ \tilde{t}_D =\sum_{n=0}^{\infty} t_{Dn} \epsilon^{n}
\end{eqnarray}
The leading term in the deformed periods are $t_0\equiv t$ and $t_{D0}\equiv t_D$. We expect the higher order contributions can be written as a linear combination of first and second derivatives of the leading term  with respect to $z_i$, and the coefficients are rational functions of $z_i$. 

Since we have two complex parameters $z_1$ and $z_2$, there are more ways to write first and second derivatives than the one-parameter case. However, as it turns out the situation is simpler than expected, and we find only the derivative $\Theta_z = z_1 \partial_{z_1}+ z_2 \partial_{z_2}$ appears in the calculations. We will find the higher order contributions $t_{2n}$ can be always written as linear combination of $\Theta_z t$ and $\Theta_z^2 t$. 

One may wonder whether the parameter  $t_{-}$ is deformed  as well. The higher order contributions would be a linear combination of  $\Theta_z t_{-}$ and $\Theta_z^2 t{-}$. It is easy to calculate actually  $\Theta_z t_{-}=0 $, so we see that  the parameter  $t_{-}$ is not deformed.  It is a nice feature of the local $\mathbb{P}^1\times \mathbb{P}^1$ model that only one linear combination of the periods $t=\frac{1}{2}(T_1+T_2)$ is deformed quantum mechanically by small $\epsilon$. Otherwise it would be rather difficult to work with the quantum differential equation (\ref{1222.qmdf}). 

We will find that the quantum differential equation (\ref{1222.qmdf}) determines the partial derivative of higher genus amplitudes $\partial_t F^{(n,0)}(t,t_{-}) $ recursively. This is actually enough to completely fix the world-sheet instanton contributions, which consists of only  positive powers of $Q_1=e^{T_1}$ and $Q_2=e^{T_2}$ and therefore can not be a function of $t_{-}$ alone. For $n\geq 2$, the differential equation  (\ref{1222.qmdf})  fixes  $ F^{(n,0)}(t,t_{-}) $ up to a constant which is the analog of the constant map contributions in Gromov-Witten theory. For $n=1$, the perturbative contribution from the B-model calculations in \cite{HK2010} is $F^{(1,0)}_{pert} =-\frac{1}{24} T_1T_2 = -\frac{1}{24}(t^2-t_{-}^2)$. We see that in this case the term $ \frac{1}{24}t_{-}^2$ is not fixed by the differential equation (\ref{1222.qmdf}). 

We solve the zero energy  quantum wave function with WKB expansion 
\begin{eqnarray}
H(x,p) \psi(x)=0,~~~ \psi(x) = \exp(\frac{1}{\epsilon} \int^xw(x) dx),~~~ w(x)=\sum_{n=0}^{\infty}w_n(x) \epsilon^n
\end{eqnarray} 
The quantum period is computed by the residue of $w(x)$ around the pole $x=0$. For the leading order $w_0(x)$, the residue only captures the power series in $t$. The higher order contribution $t_n$ for $n\geq 1$ is given exactly by the residue of $w_n(x)$ around $x=0$. We find the exact formulae for the first few higher even order non-vanishing contributions to the quantum period 
\begin{eqnarray}
t_2 &=& -\frac{z_1+z_2}{6} \Theta_z t + \frac{1-4z_1-4z_2}{12} \Theta_z^2 t \nonumber \\
t_4 &=& \frac{1}{360\Delta^2} \{ 2[  z_1^2 (1 - 4 z_1)^3 + z_2^2 (1 - 4 z_2)^3+4z_1z_2(8 - 37 z_1  - 37 z_2 - 328 z_1^2 + 1528 z_1 z_2  \nonumber \\ && - 328 z_2^2    + 1392 z_1^3 
-  1376 z_1^2 z_2  - 1376 z_1 z_2^2 + 1392 z_2^3)]\Theta_z t +[  -z_1 (1 - 4 z_1)^4 - z_2 (1 - 4 z_2)^4 \nonumber \\ && +4z_1z_2(69 - 192 z_1 - 192 z_2 - 1712 z_1^2 + 6880 z_1 z_2 - 1712 z_2^2 + 5568 z_1^3   - 
 5504 z_1^2 z_2 \nonumber \\ &&  - 5504 z_1 z_2^2 + 5568 z_2^3 )]\Theta_z^2 t \}
\end{eqnarray} 
The dual period satisfies the same equations with $t$ replaced by  $t_D$.    

In \cite{HK2010} we find the formulae for higher genus amplitudes by holomorphic anomaly equations and the gap boundary conditions. Some examples of the formulae in the Nekrasov-Shatashvili limit are
\begin{eqnarray} \label{1223.higher}
F^{(1,0)} &=&  \frac{1}{24}\log(\frac{\Delta}{z_1z_2}), \nonumber \\
 F^{(2,0)} &=&
\frac{S}{288 z_1^2 \Delta^2}   \left(16 z_1^2+32  z_1z_2+16 z_2^2 -8 z_1-8 z_2+1\right)+\frac{1}{2880\Delta^2}( -512 z_1^4    \nonumber \\&& +9216z_1^3 z_2 -17408  z_1^2z_2^2 +9216 z_1 z_2^3-512 z_2^4 +704 z_1^3 +2880
   z_1^2  z_2 +2880  z_1 z_2^2      \nonumber \\&&    +704 z_2^3-336 z_1^2 -1568 z_1z_2  -336 z_2^2+68 z_1 +68 z_2-5)
\end{eqnarray}
Here  as it turns out that the BCOV propagators $S^{z_iz_j}$ ($i,j=1,2$) are not independent and all propagators are related to one which we can choose as $S\equiv S^{z_1z_1}$. The derivatives of the higher genus formulae can then be written as rational functions of five independent generators $z_1$, $z_2$, $\partial_{z_1} t$, $\partial_{z_2} t$ and  $S$, whose derivatives  are rational functions of themselves. The calculations of the derivative rules are provided by the formulae in \cite{HK2010}. We omit the details here as it is similar to the local $\mathbb{P}^2$ model in the previous subsection. 

With the derivative rules for the five generators, we expand the quantum differential equation (\ref{1222.qmdf}) for small $\epsilon$ and check our higher genus formulae (\ref{1223.higher}) for $F^{(n,0)}$ ($n\geq 1$) exactly  satisfy the equations  at each order of the small $\epsilon$ expansion.

\section{Conclusion}

There are some questions for further study. The equation for the deformed dual period $\frac{\partial \mathcal{F}(\tilde{a})}{\partial \tilde{a}} = \tilde{a}_D $ should be derived more carefully, e.g. from the saddle point analysis for the Nekrasov function in Seiberg-Witten theory and from the refined topological vertex in toric Calabi-Yau models. 

We have provided a derivation of the holomorphic anomaly equations in the  Nekrasov-Shatashvili limit from the differential equations for the deformed dual period.  It would be nice to also derive the gap boundary conditions  from these differential equations.

\vspace{0.2in} {\leftline {\bf Acknowledgments}}

We thanks Amir-Kian Kashani-Poor and Albrecht Klemm for discussions and correspondences.

\end{document}